\documentclass[a4paper,11pt]{article}
\pdfoutput=1 

\usepackage{jcappub} 

\usepackage[T1]{fontenc} 

\newcommand{\orcid}[1]{\href{https://orcid.org/#1}{\,\includegraphics[width=8px]{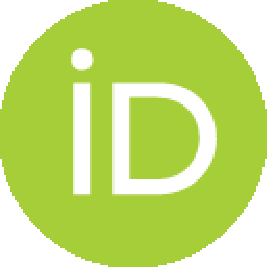}}}

\title{\boldmath A new diagnostic for the null test of dynamical dark energy in light of DESI 2024 and other BAO data}

\author[a,1]{Bikash R. Dinda\orcid{0000-0001-5432-667X}\note{Corresponding author.}}

\affiliation[a]{Department of Physics $\&$ Astronomy, University of the Western Cape, Cape Town, 7535, South Africa.}

\emailAdd{bikashrdinda@gmail.com}

\abstract{We introduce a new diagnostic for the null tests of dynamical dark energy alongside two other combined equivalent diagnostics. These diagnostics are useful, especially when we include anisotropic baryon acoustic oscillation (BAO) data in an analysis, to quantify the deviations from the standard $\Lambda$CDM model. We also consider another diagnostic for isotropic BAO observations. These null tests are independent of any late-time dark energy model or parametrization. With these diagnostics, we study the evidence for dynamical dark energy in light of Dark Energy Spectroscopic Instrument (DESI) 2024 data combined with cosmic microwave background (CMB) observations of the Planck 2018 mission and local $H_0$ measurements. We find no strong evidence for dynamical dark energy. The exclusion of the individual deviations at the effective redshift 0.51 of the DESI 2024 data makes the evidence even weaker. We get nearly similar results for other non-DESI BAO data. Both for DESI 2024 and other non-DESI BAO data, the evidence is almost independent of early-time physics. The evidence corresponding to the SHOES value of $H_0$ is higher than the corresponding tRGB value of $H_0$ for all combinations of data, but still not strong enough to reject the flat $\Lambda$CDM model.}

\begin{document}
\maketitle
\flushbottom

\section{Introduction}

Ever since the discovery of the late time cosmic acceleration from the type Ia supernova observations \citep{SupernovaCosmologyProject:1997zqe,SupernovaSearchTeam:1998fmf,SupernovaCosmologyProject:1998vns,2011NatPh...7Q.833W}, this phenomenon has been confirmed by several other observations such as cosmic microwave background (CMB) observations \citep{Planck:2013pxb,Planck:2015fie,Planck:2018vyg}, baryon acoustic oscillation (BAO) observations \citep{BOSS:2016wmc,eBOSS:2020yzd,Hou:2020rse}, and cosmic chronometers (CC) observations for the Hubble parameter \citep{Jimenez:2001gg,Pinho:2018unz,Cao:2023eja} etc.

Arguably, the most accepted explanation for the late time cosmic acceleration is the introduction of an exotic matter, called dark energy, as a possible constituent of the Universe and most importantly the dominant one in the late time era. Unlike any other matter, due to the large negative pressure of the dark energy, it has an effective repulsive gravity which in turn causes the late time cosmic acceleration of the Universe \citep{Peebles:2002gy,Copeland:2006wr,Yoo:2012ug,Lonappan:2017lzt,Dinda:2017swh,Dinda:2018uwm,Bamba:2012cp,Dinda:2024xla}.

Since the introduction of the concept of dark energy, model building and phenomenological studies of the nature of dark energy have become crucial for understanding the evolutionary history of the Universe. Regarding this, the most popular model of dark energy is the $\Lambda$CDM model. In this model, a cosmological constant $\Lambda$ is considered to be the candidate for the dark energy and the equation of state of this $\Lambda$ is $-1$ which causes the late time expansion of the Universe to be accelerating \citep{Carroll:2000fy}.

The $\Lambda$CDM model stood out to be the most successful model to explain the majority of the cosmological observations, mentioned above. This is the reason it became the most popular model for the evolution of the Universe. It is sometimes called the standard model of cosmology (with the inclusion of inflation at a very early time era). Despite its huge success, there are some serious issues with this model. For example, theoretically, this model possesses the fine-tuning and cosmic coincidence problems \citep{Zlatev:1998tr,Sahni:1999gb,Velten:2014nra,Malquarti:2003hn}. From the observational point of view, there are issues like the Hubble tension \citep{DiValentino:2021izs,Krishnan:2021dyb,Vagnozzi:2019ezj,Dinda:2021ffa}, in which there is a discrepancy in the observed values of the Hubble parameter between the early time (e.g. CMB \citep{Planck:2018vyg}) and the late time (e.g. SHOES \citep{Riess:2020fzl}) observations.

Because of these shortcomings of the $\Lambda$CDM model, it is thus important to consider dynamical dark energy models to study the nature of the dark energy and to check evidence of these alternative models in light of recent and upcoming observations. Note that, in the $\Lambda$CDM model, the energy density of the dark energy is constant in cosmic time. Hence, any model with evolving energy density of the dark energy is considered to be the dynamical dark energy models such as quintessence \citep{Dinda:2016ibo}, k-essence \citep{Dinda:2023mad}, and Chevallier-Polarski-Linder (CPL) \citep{Chevallier:2000qy,Linder:2002et} models.

Recently, the Dark Energy Spectroscopic Instrument (DESI) 2024 Data Release 1 (DR1) BAO observations (with and without the addition of CMB \citep{Planck:2018vyg} and type Ia supernova observations \citep{Brout:2022vxf,Rubin:2023ovl,DES:2024tys} both separately and jointly) have found more than 2$\sigma$ evidence for the dynamical dark energy considering the CPL model (most commonly known as the $w_0w_a$CDM model) \citep{DESI:2024mwx}. Similar results with similar data are presented in a different analysis in Wang (2024) \citep{Wang:2024rjd}. Interestingly, with the $w_0w_a$CDM model, similar kind of evidence for dynamical dark energy has been reported with other BAO data in Park et al. (2024) \citep{Park:2024jns}. Besides these, the DESI 2024 data along with other data like CMB and the type Ia supernova have been considered to constrain different model parameters in different models \citep{Tada:2024znt,Wang:2024hks,Carloni:2024zpl,Berghaus:2024kra,Giare:2024smz,Qu:2024lpx,Wang:2024dka,Yang:2024kdo,Yin:2024hba,Escamilla-Rivera:2024sae,Huang:2024qno,Wang:2024pui}. Also, note that the $\Lambda$CDM model has been revisited with the DESI 2024 data in a different approach through the estimation of the present value of matter energy density parameter in Colg\^ain et al. (2024) \citep{Colgain:2024xqj}.

Another interesting fact is that the same DESI 2024 data (with or without the same other data, mentioned earlier) shows that there is no evidence for dynamical dark energy when considering the $w$CDM model. Note that, while both $w$CDM and $w_0w_a$CDM models correspond to the dynamical dark energy, the equation of state of dark energy is constant in the $w$CDM model but evolving in $w_0w_a$CDM model. In another DESI paper by Calderon et al. (2024) \citep{Calderon:2024uwn}, evidence for dynamical dark energy has been studied through crossing statistics and the results are different between crossing statistics applied to equation of state and the (normalized) energy density of dark energy (compare figures 1 and 3 in Calderon et al. (2024) \citep{Calderon:2024uwn}). In another study, with similar data, Luongo $\&$ Muccino (2024) \citep{Luongo:2024fww} has reported that the $w$CDM model is favored over the CPL model through a model-independent cosmographic approach. Thus, the evidence for dynamical dark energy may be biased depending on different models of dark energy \citep{Linder:2006xb,Wolf:2023uno}. Also, the priors on the dark energy parameters (see Cort\^es $\&$ Liddle (2024) \citep{Cortes:2024lgw}) and the degeneracies (see Shlivko $\&$ Steinhardt (2024) \citep{Shlivko:2024llw}) may affect the evidence if we consider a particular dark energy model or parametrization. Thus, in this regard, any dark energy model-independent (and for the betterment non-parametric too) approach will be useful to firmly test the evidence for dynamical dark energy.

In this study, we perform dark energy model-independent null tests of the evidence for dynamical dark energy by introducing a new diagnostic approach to study the deviation from the standard $\Lambda$CDM model. Because this approach is independent of any dark energy model, there is no issue of bias in the results. Also, the results are independent of any priors on the dark energy parameters, since there are no such parameters involved in this null test. Thus, this kind of null test is crucial for the evidence of dynamical dark energy. The diagnostics, introduced in this analysis, are more useful compared to other existing diagnostics like the Om diagnostic \citep{Sahni:2008xx,Myrzakulov:2023rxp}, particularly when we consider BAO observations because these diagnostics take into account both the parallel and perpendicular anisotropic BAO measurements to the line of sight and their correlations too.

This paper is organized as follows: In Sec.~\ref{sec-basic}, we mention basic equations used in this analysis; In Sec.~\ref{sec-diagnostic}, we discuss four different diagnostics for the null tests of dynamical dark energy in light of BAO observations; in Sec.~\ref{sec-data}, we briefly discuss different cosmological data, considered in this analysis, including the recent DESI 2024 data; in Sec.~\ref{sec-result}, we present the results for the evidence of dynamical dark energy corresponding to different data combinations; and finally, we conclude this study in Sec.~\ref{sec-conclusion}. We have also included some relevant details in the appendices. In Appendix~\ref{sec-los_Lcdm}, we have provided the derivation for the analytical expression of the comoving distance as a function of redshift for the $\Lambda$CDM model. In Appendix~\ref{sec-appx_sound_horzon_early_standard}, we have provided an analytical approximate expression for the sound horizon applicable at higher redshift with the standard early-time cosmological physics, and in Appendix~\ref{sec-appx_photon_decoupling_z_standard}, we have provided a corresponding approximate expression of the redshift of photon decoupling. Also, in Appendix~\ref{sec-appx_baryon_drag_z_standard}, we provide an approximate expression of baryon drag redshift with the same approximation. In Appendix~\ref{sec-cmb_dist_prior_again}, we mention the CMB distance priors, especially the covariance matrices both for standard and non-standard early physics. Finally, in Appendix~\ref{sec-dark_energy_parametrizations}, we have briefly discussed the $w$CDM and the $w_0w_a$CDM parametrizations.

\section{Basic equations}
\label{sec-basic}

We consider flat Friedmann-Lema\^itre-Robertson-Walker (FLRW) metric for the background evolution of the Universe. For this case, the transverse comoving distance $D_M$ is the same as the line of sight comoving distance and it is defined as \citep{Hogg:1999ad}

\begin{equation}
D_M(z) = \frac{c}{H_0} \int_{0}^{z} \frac{d\tilde{z}}{E(\tilde{z})},
\label{eq:defn_DM}
\end{equation}

\noindent
where $z$ (also $\tilde{z}$) is the redshift, $c$ is the speed of light in vacuum, $H_0$ is the present value of the Hubble parameter, and $E$ is the normalized Hubble parameter.

For a flat $\Lambda$CDM model, for the late-time evolution, the normalized Hubble parameter (denoted as $E^{\rm \Lambda CDM}$) is given as

\begin{equation}
E^{\rm \Lambda CDM}(z) = \sqrt{\Omega_{\rm m0}(1+z)^3+1-\Omega_{\rm m0}} ~ ,
\label{eq:E_LCDM_late}
\end{equation}

\noindent
where $\Omega_{\rm m0}$ is the present value of the matter energy density parameter. In Eq.~\eqref{eq:E_LCDM_late}, we have neglected the contribution from the radiation counterpart because we are using this equation for the late-time evolution only.

Putting Eq.~\eqref{eq:E_LCDM_late} in Eq.~\eqref{eq:defn_DM}, and performing the integration analytically, we get a corresponding expression for the comoving distance given as \citep{Dinda:2021ffa}

\begin{equation}
D_M^{\rm \Lambda CDM} (z) = \frac{c F(z)}{H_0 \sqrt{1-\Omega _{\text{m0}}}} ,
\label{eq:DM_LCDM_late}
\end{equation}

\noindent
where $F$ is given as

\begin{equation}
F (z) = (1+z) \, _2F_1 \left[ 1/3,1/2;4/3;-\alpha(1+z)^3 \right] - \, _2F_1 \left[ 1/3,1/2;4/3;-\alpha \right] ,
\label{eq:los_Lcdm_main_F}
\end{equation}

\noindent
where $_2F_1$ represents the standard hypergeometric function and $\alpha$ is given as

\begin{equation}
\alpha = \frac{\Omega_{\rm m0}}{1-\Omega_{\rm m0}} .
\label{eq:defn_alpha}
\end{equation}

\noindent
We have included the derivation of Eq.~\eqref{eq:DM_LCDM_late} for $D_M^{\rm \Lambda CDM}$ in Appendix~\ref{sec-los_Lcdm}.

Computation of $\Omega_{\rm m0}$ is not very trivial, because it is a derived parameter. Hence late-time dark energy model independent computation of $\Omega_{\rm m0}$ is non-trivial. However, it can be indirectly computed from the combined parameter $\Omega_{\rm m0}h^2$ and $h$, where $h$ is related to $H_0$ as

\begin{equation}
H_0 = 100~h~{\rm Km}~{\rm s}^{-1}~{\rm Mpc}^{-1},
\label{eq:defn_h}
\end{equation}

\noindent
We will see later that $\Omega_{\rm m0}h^2$ can be computed from observations like CMB almost independent of any late-time dark energy model (it depends mainly on early-time physics). The parameter $h$ can be obtained from any local measurement of $H_0$, mentioned later. So, the computation of $\Omega_{\rm m0}$ or $\alpha$ (defined in Eq.~\eqref{eq:defn_alpha}) depends on $\Omega_{\rm m0}h^2$ and $h$. So, for a late-time dark energy model-independent analysis, it is useful to rewrite the expression of $\Omega_{\rm m0}$ or $\alpha$ w.r.t $\Omega_{\rm m0}h^2$ and $h$. Eq.~\eqref{eq:defn_alpha} can be alternatively written as

\begin{equation}
\alpha = \frac{\omega_{\rm m0}}{h^2-\omega_{\rm m0}} ,
\label{eq:defn_alpha_again}
\end{equation}

\noindent
and $\omega_{\rm m0}$ is given as

\begin{equation}
\omega_{\rm m0} = \Omega_{\rm m0} h^2.
\label{eq:defn_wm0}
\end{equation}

A quantity $D_H$ is defined as

\begin{equation}
D_H(z) = \frac{c}{H(z)} = \frac{c}{H_0 E(z)} ,
\label{eq:defn_DH}
\end{equation}

\noindent
where $H$ is the Hubble parameter. Putting Eq.~\eqref{eq:E_LCDM_late} in Eq.~\eqref{eq:defn_DH}, we get the corresponding expression for $D_H$ for the $\Lambda$CDM model at late times given as

\begin{equation}
D_H^{\rm \Lambda CDM} (z) = \frac{c}{H_0 \sqrt{\Omega_{\rm m0}(1+z)^3+1-\Omega_{\rm m0}}} .
\label{eq:DH_LCDM_late}
\end{equation}

Another quantity $D_V$ is defined as

\begin{equation}
D_V(z) = \left[zD_M^2(z)D_H(z)\right]^{1/3} .
\label{eq:defn_DV}
\end{equation}

\noindent
Putting Eqs.~\eqref{eq:DM_LCDM_late} and~\eqref{eq:DH_LCDM_late} in Eq.~\eqref{eq:defn_DV}, we get the corresponding expression for $D_V$ for the $\Lambda$CDM model at late times given as

\begin{equation}
D_V^{\Lambda CDM} (z) = \frac{c}{H_0} \left[ \frac{zF^2(z)}{(1-\Omega_{\rm m0})\sqrt{\Omega_{\rm m0}(1+z)^3+1-\Omega_{\rm m0}}} \right]^{1/3} .
\label{eq:DV_LCDM_late}
\end{equation}

\subsection{BAO observables}
\label{sec-bao_observables}

The baryon acoustic oscillation (BAO) observations are quantified with three quantities given as

\begin{equation}
\tilde{D}_M (z) = \frac{D_M(z)}{r_d} , ~~~~~ \tilde{D}_H (z) = \frac{D_H(z)}{r_d} , ~~~~~ \tilde{D}_V (z) = \frac{D_V(z)}{r_d} ,
\label{eq:BAO_quantities}
\end{equation}
where $r_d$ is the sound horizon at the baryon drag epoch i.e. the distance that sound travelled from Big Bang to the baryon drag epoch. $D_M$ is the transverse (perpendicular to the observers' line of sight direction) comoving distance. $D_H$ is another distance variable parallel to the line of sight. The BAO observations do not observe these distances directly but the ratios $\tilde{D}_M$ and $\tilde{D}_H$, defined in Eq.~\eqref{eq:BAO_quantities}. These are distances normalized to the sound horizon at the baryon drag epoch $r_d$. $D_V$ is the angle averaged distance and thus the observations related to it are isotropic, whereas observations related to variables $D_M$ and $D_H$ are anisotropic. Similar to $D_M$ and $D_H$, BAO observations do not directly observe $D_V$ but rather $D_V/r_d$.

\section{Diagnostics in light of BAO}
\label{sec-diagnostic}

\subsection{The diagnostic $A_1$}
\label{sec-diagnostic_A1}

We define a diagnostic $A_1$ as $\tilde{D}_M/\tilde{D}_M^{\rm \Lambda CDM}$ which has the corresponding expression given as

\begin{equation}
A_1 (z) = \frac{r_d H_0 \sqrt{1-\Omega_{\rm m0}} \tilde{D}_M(z)}{c F(z)} \approx \frac{r_d \sqrt{h^2-\omega_{\rm m0}} \tilde{D}_M(z)}{3000 ~ \rm{Mpc} ~ F(z)} .
\label{eq:diagnostic_A1}
\end{equation}

\noindent
In the second (approximate) equality, we have used the fact that

\begin{equation}
\frac{H_0}{c} \simeq \frac{h}{3000 ~ {\rm Mpc}} .
\label{eq:H0_by_c}
\end{equation}

\noindent
In Eq.~\eqref{eq:diagnostic_A1}, $A_1$ is defined in such a way that its corresponding value for the $\Lambda$CDM model is $1$ i.e.

\begin{equation}
A_1^{\rm \Lambda CDM} (z) = 1 .
\label{eq:A1_LCDM_late}
\end{equation}

\subsection{The diagnostic $A_2$}
\label{sec-diagnostic_A2}

We define a second diagnostic $A_2$ as $\tilde{D}_H/\tilde{D}_H^{\rm \Lambda CDM}$ which has the corresponding expression given as

\begin{equation}
A_2 (z) = \frac{r_d H_0 \sqrt{\Omega_{\rm m0}(1+z)^3+1-\Omega_{\rm m0}} \tilde{D}_H(z)}{c} \approx \frac{r_d \sqrt{\omega_{\rm m0}(1+z)^3+h^2-\omega_{\rm m0}} \tilde{D}_H(z)}{3000~{\rm Mpc}} .
\label{eq:diagnostic_A2}
\end{equation}

\noindent
In the above equation, $A_2$ is defined in such a way that its corresponding value for the $\Lambda$CDM model is $1$ i.e.

\begin{equation}
A_2^{\rm \Lambda CDM} (z) = 1 .
\label{eq:A2_LCDM_late}
\end{equation}

\subsection{The new diagnostic: $B$}
\label{sec-new_diagnostic_B}

We define a quantity $F_{\rm AP}$ given as \citep{DESI:2024mwx,Colgain:2024xqj,Colgain:2022nlb}

\begin{equation}
F_{\rm AP}(z) = \frac{\tilde{D}_M(z)}{\tilde{D}_H(z)}.
\label{eq:defn_FAP}
\end{equation}

\noindent
The reason to define the quantity $F_{\rm AP}$ in this way is as follows. The BAO observations provide data of $\frac{D_M}{r_d}$ and $\frac{D_H}{r_d}$, but not directly $D_M$ and $D_H$. So, when we divide $\tilde{D}_M$ by $\tilde{D}_H$, the ratio becomes independent of $r_d$. In general case, we get the expression of $F_{\rm AP}$, by putting Eqs.~\eqref{eq:defn_DM} and~\eqref{eq:defn_DH} in Eq.~\eqref{eq:defn_FAP}, given as

\begin{equation}
F_{\rm AP}(z) = \frac{\tilde{D}_M(z)}{\tilde{D}_H(z)} = \frac{D_M(z)}{D_H(z)} = E(z) \int_{0}^{z} \frac{d\tilde{z}}{E(\tilde{z})}.
\label{eq:general_FAP}
\end{equation}

\noindent
Putting Eqs.~\eqref{eq:DM_LCDM_late} and~\eqref{eq:DH_LCDM_late} in Eq.~\eqref{eq:defn_FAP} (or in Eq.~\eqref{eq:general_FAP}), we get the corresponding expression for $F_{\rm AP}$ for the $\Lambda$CDM model at late times given as

\begin{equation}
F_{\rm AP}^{\rm \Lambda CDM} (z) = F(z) \sqrt{\alpha(1+z)^3+1} .
\label{eq:FAP_LCDM_late}
\end{equation}

We now define a third diagnostic $B$ as $B=F_{\rm AP}/F_{\rm AP}^{\rm \Lambda CDM}$ which has the corresponding expression given as

\begin{equation}
B(z) = \frac{F_{\rm AP}}{F(z) \sqrt{\alpha(1+z)^3+1}} .
\label{eq:defn_B}
\end{equation}

\noindent
So, in the above equation, $B$ is defined in such a way that its corresponding value for the $\Lambda$CDM model is $1$ i.e.

\begin{equation}
B^{\rm \Lambda CDM} (z) = 1 .
\label{eq:B_LCDM_late}
\end{equation}

Eq.~\eqref{eq:defn_B} can be useful for the dark energy model-independent null test for the evidence of dynamical dark energy, especially, when BAO data is considered. Given the values of $\alpha$ and $F_{\rm AP}(z)$ from any observations, we can compute $B(z)$ from Eq.~\eqref{eq:defn_B} and if this value deviates from unity at late times, we get evidence for dynamical dark energy.

\subsection{The diagnostic $A_3$}
\label{sec-diagnostic_A3}

We define a fourth diagnostic $A_3$ as $\tilde{D}_V/\tilde{D}_V^{\Lambda CDM}$ which has the corresponding expression given as

\begin{eqnarray}
A_3 (z) &=& \frac{r_dH_0\tilde{D}_V}{c} \left[ \frac{(1-\Omega_{\rm m0})\sqrt{\Omega_{\rm m0}(1+z)^3+1-\Omega_{\rm m0}}}{zF^2(z)} \right]^{1/3} , \nonumber\\
&\approx& \frac{r_d\tilde{D}_V}{3000~{\rm Mpc}} \left[ \frac{(h^2-\omega_{\rm m0})\sqrt{\omega_{\rm m0}(1+z)^3+h^2-\omega_{\rm m0}}}{zF^2(z)} . \right]^{1/3}
\label{eq:diagnostic_A3}
\end{eqnarray}

\noindent
In the above equation, $A_3$ is defined in such a way that its corresponding value for the $\Lambda$CDM model is $1$ i.e.

\begin{equation}
A_3^{\Lambda CDM} (z) = 1 .
\label{eq:A3_LCDM_late}
\end{equation}

The advantage of all these diagnostics (when combined accordingly) is that these are more useful compared to other existing diagnostics such as the Om diagnostic \citep{Sahni:2008xx,Myrzakulov:2023rxp} (most of the existing diagnostics take into account only the Hubble parameter related data such as $H$ and $D_H$) when we consider BAO data because these take into account anisotropic BAO data in both transverse and line of sight directions and isotropic BAO data too. This advantage is not present in most of the diagnostics, in which one can relate the diagnostic variables to only $\tilde{D}_H$, but not to $\tilde{D}_M$, because the inclusion of $\tilde{D}_M$ may require its derivative (or alternatively integration in $\tilde{D}_H$) which makes the analysis complicated. It would be even more complicated if one wants to include simultaneously both $\tilde{D}_M$ and $\tilde{D}_H$ and their correlations too.

\section{Observational data}
\label{sec-data}

\subsection{BAO data}
\label{sec-bao_data}

\subsubsection{DESI 2024 BAO}
\label{sec-desi_2024}

We consider Dark Energy Spectroscopic Instrument (DESI) 2024 BAO observations \citep{DESI:2024mwx} as the main data in our analysis. The anisotropic BAO data contain five main samples which are the Luminous Red Galaxy samples (LRG) at two effective redshifts $z_{\rm eff}=0.51$ and $z_{\rm eff}=0.71$ \citep{DESI:2022gle}, combinations of LRG and the Emission Line Galaxy sample (ELG) at effective redshift $z_{\rm eff}=0.93$ \citep{DESI:2024mwx}, the solo ELG sample at effective redshift $z_{\rm eff}=1.32$ \citep{Raichoor:2022jab}, and the Lyman-$\alpha$ forest sample (Ly-$\alpha$) at effective redshift $z_{\rm eff}=2.33$ \citep{DESI:2024mwx}. We quote the mean and standard deviation values of these two variables in Table~\ref{table:DESI24_DM_DH_tilde} corresponding to the DESI 2024 observations. We also consider two isotropic BAO data. One is at $z_{\rm eff}=0.295$ corresponding to the Bright Galaxy Sample (BGS) \citep{DESI:2024mwx}. The second one is at $z_{\rm eff}=1.491$ corresponding to the Quasar Sample (QSO) \citep{Chaussidon:2022pqg}. These are also listed in Table~\ref{table:DESI24_DM_DH_tilde}. We call these data 'DESI24' throughout this analysis.

\begin{table*}
\begin{center}
DESI 2024: Anisotropic BAO data \\
\begin{tabular}{|c|c|c|c|c|c|}
\hline
tracer (DESI24) &$z_{\rm eff}$ & $\tilde{D}_M \pm \Delta \tilde{D}_M$ & $\tilde{D}_H \pm \Delta \tilde{D}_H$ & $r_{\rm MH}$ & Refs. \\
\hline
LRG & 0.51 & 13.62 $\pm$ 0.25 & 20.98 $\pm$ 0.61 & -0.445 & \citep{DESI:2022gle} \\
LRG & 0.71 & 16.85 $\pm$ 0.32 & 20.08 $\pm$ 0.60 & -0.420 & \citep{DESI:2022gle} \\
LRG+ELG & 0.93 & 21.71 $\pm$ 0.28 & 17.88 $\pm$ 0.35 & -0.389 & \citep{DESI:2024mwx} \\
ELG & 1.32 & 27.79 $\pm$ 0.69 & 13.82 $\pm$ 0.42 & -0.444 & \citep{Raichoor:2022jab} \\
Ly-$\alpha$ & 2.33 & 39.71 $\pm$ 0.94 & 8.52 $\pm$ 0.17 & -0.477 & \citep{DESI:2024mwx} \\
\hline
\end{tabular} \\
DESI 2024: Isotropic BAO data \\
\begin{tabular}{|c|c|c|c|}
\hline
tracer (DESI24) &$z_{\rm eff}$ & $\tilde{D}_V \pm \Delta \tilde{D}_V$ & Refs. \\
\hline
BGS & 0.295 & 7.93 $\pm$ 0.15 & \citep{DESI:2024mwx} \\
QSO & 1.491 & 26.07 $\pm$ 0.67 & \citep{Chaussidon:2022pqg} \\
\hline
\end{tabular}
\end{center}
\caption{
The observed values of $\tilde{D}_M$, $\tilde{D}_H$, corresponding 1$\sigma$ errors, the correlations between them at five different effective redshifts, and $\tilde{D}_V$ with associated 1$\sigma$ errors at two different effective redshifts obtained directly from the DESI 2024 data (denoted as 'DESI24') \citep{DESI:2024mwx}.
}
\label{table:DESI24_DM_DH_tilde}
\end{table*}

The $\Delta$ notation in front of any quantity corresponds to the standard deviation (1$\sigma$ error) of that quantity throughout this paper. Note that, throughout the entire analysis, we use the notation '+' between any names of observations to convey that the observations have been combined for any analysis. We have also quoted the correlations between $\tilde{D}_M$ and $\tilde{D}_H$ with the notation $r_{\rm MH}$ defined as

\begin{equation}
r_{\rm MH} = r(\tilde{D}_M,\tilde{D}_H) = \frac{{\rm Cov} [ \tilde{D}_M,\tilde{D}_H ]}{\Delta \tilde{D}_M ~ \Delta \tilde{D}_H} ,
\label{eq:defn_rMH}
\end{equation}

\noindent
where ${\rm Cov} [ \tilde{D}_M,\tilde{D}_H ]$ denotes the covariances between $\tilde{D}_M$ and $\tilde{D}_H$ and notation $r(X,Y)$ denotes the correlation coefficient or normalized covariance between any two quantities $X$ and $Y$.

\begin{table}
\begin{center}
\begin{tabular}{|c|c|c|c|}
\hline
tracer (DESI24) &$z_{\rm eff}$ & $F_{\rm AP}\pm \Delta F_{\rm AP}$ & Refs. \\
\hline
LRG & 0.51 & 0.649 $\pm$ 0.026 & \citep{DESI:2022gle} \\
LRG & 0.71 & 0.839 $\pm$ 0.035 & \citep{DESI:2022gle} \\
LRG+ELG & 0.93 & 1.214 $\pm$ 0.033 & \citep{DESI:2024mwx} \\
ELG & 1.32 & 2.011 $\pm$ 0.095 & \citep{Raichoor:2022jab} \\
Ly-$\alpha$ & 2.33 & 4.661 $\pm$ 0.175 & \citep{DESI:2024mwx} \\
\hline
\end{tabular}
\end{center}
\caption{
The measurements of $F_{\rm AP}$ with 1$\sigma$ error bars at five different effective redshifts corresponding to the DESI 2024 data for anisotropic BAO observations.
}
\label{table:DESI_24_FAP_data}
\end{table}

\begin{table*}
\begin{center}
\begin{tabular}{|c|c|c|c|c|c|}
\hline
tracer (non-DESI BAO) &$z_{\rm eff}$ & $\tilde{D}_M \pm \Delta \tilde{D}_M$ & $\tilde{D}_H \pm \Delta \tilde{D}_H$ & $r_{\rm MH}$ & Refs. \\
\hline
LRG (BOSS DR12) & 0.38 & 10.234 $\pm$ 0.151 & 24.981 $\pm$ 0.582 & -0.228 & \citep{BOSS:2016wmc} \\
LRG (BOSS DR12) & 0.51 & 13.366 $\pm$ 0.179 & 22.317 $\pm$ 0.482 & -0.233 & \citep{BOSS:2016wmc} \\
LRG (eBOSS DR16) & 0.698 & 17.858 $\pm$ 0.302 & 19.326 $\pm$ 0.469 & -0.239 & \citep{eBOSS:2020yzd} \\
QSO (eBOSS DR16) & 1.48 & 30.688 $\pm$ 0.789 & 13.261 $\pm$ 0.469 & 0.039 & \citep{eBOSS:2020gbb} \\
Ly-$\alpha$ QSO (eBOSS DR16) & 2.334 & 37.5 $\pm$ 1.2 & 8.99 $\pm$ 0.19 & -0.45 & \citep{eBOSS:2020tmo} \\
\hline
\end{tabular}
\end{center}
\caption{
The observed values of $\tilde{D}_M$, $\tilde{D}_H$, corresponding 1$\sigma$ errors, and the correlations between them at five different effective redshifts obtained directly from the other BAO data corresponding to anisotropic BAO observations \citep{eBOSS:2020yzd} (denoted as 'non-DESI BAO').
}
\label{table:Other_BAO_DM_DH_tilde}
\end{table*}

\begin{table}
\begin{center}
\begin{tabular}{|c|c|c|c|}
\hline
tracer (non-DESI BAO) &$z_{\rm eff}$ & $F_{\rm AP}\pm \Delta F_{\rm AP}$ & Refs. \\
\hline
LRG (BOSS DR12) & 0.38 & 0.410 $\pm$ 0.012 & \citep{BOSS:2016wmc} \\
LRG (BOSS DR12) & 0.51 & 0.599 $\pm$ 0.017 & \citep{BOSS:2016wmc} \\
LRG (eBOSS DR16) & 0.698 & 0.924 $\pm$ 0.030 & \citep{eBOSS:2020yzd} \\
QSO (eBOSS DR16) & 1.48 & 2.314 $\pm$ 0.099 & \citep{eBOSS:2020gbb} \\
Ly-$\alpha$ QSO (eBOSS DR16) & 2.334 & 4.171 $\pm$ 0.190 & \citep{eBOSS:2020tmo} \\
\hline
\end{tabular}
\end{center}
\caption{
The measurements of $F_{\rm AP}$ with 1$\sigma$ error bars at five different effective redshifts for other non-DESI BAO data corresponding to anisotropic BAO observations \citep{eBOSS:2020yzd}.
}
\label{table:Other_BAO_FAP_data}
\end{table}

\begin{figure}
\includegraphics[height=120pt,width=0.49\textwidth]{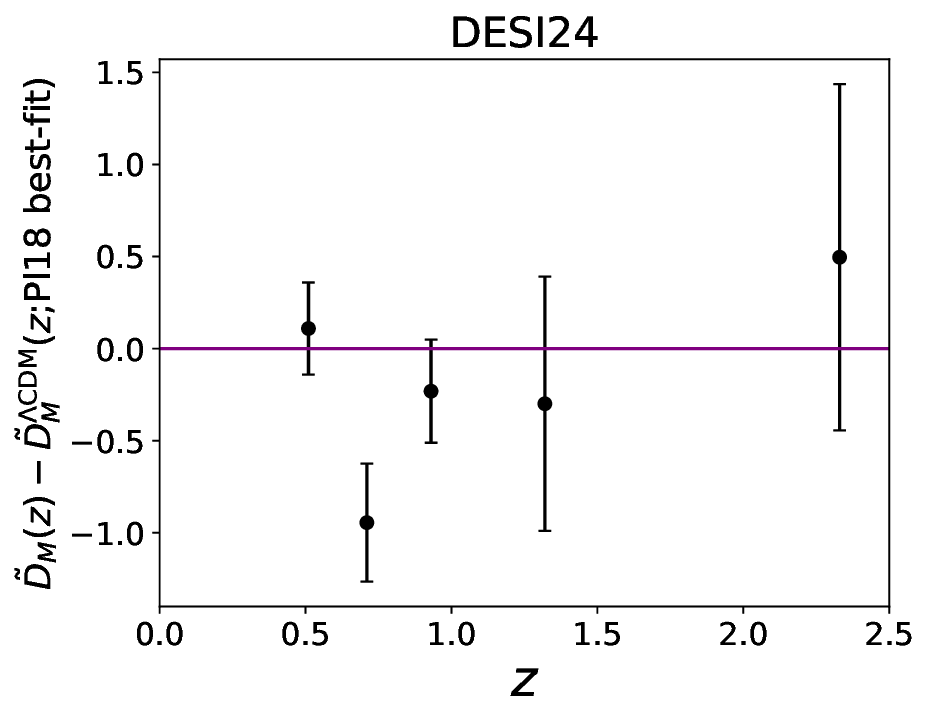}
\includegraphics[height=120pt,width=0.49\textwidth]{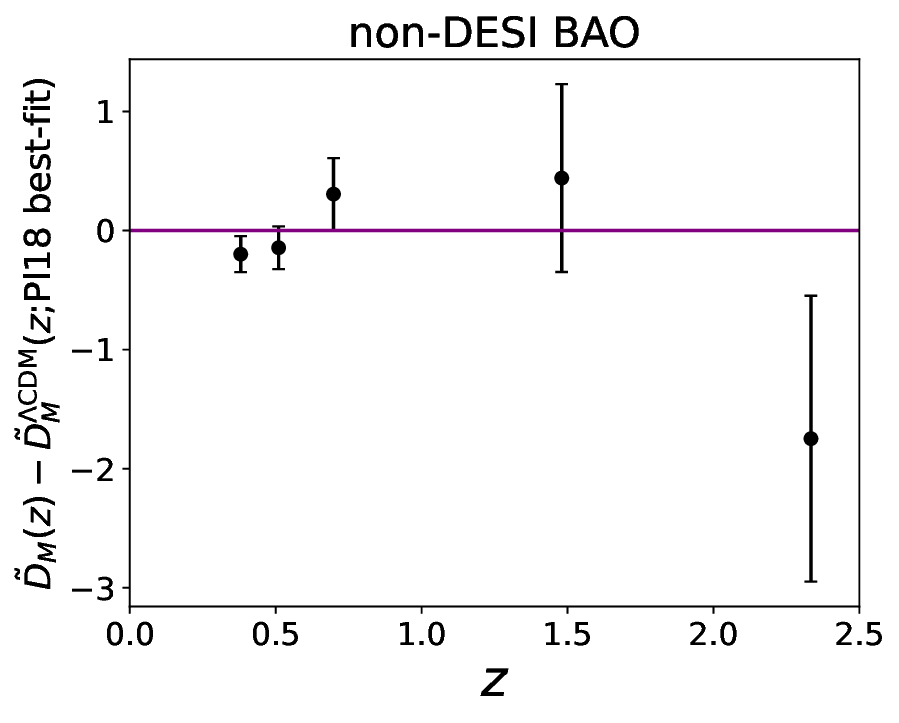} \\
\includegraphics[height=120pt,width=0.49\textwidth]{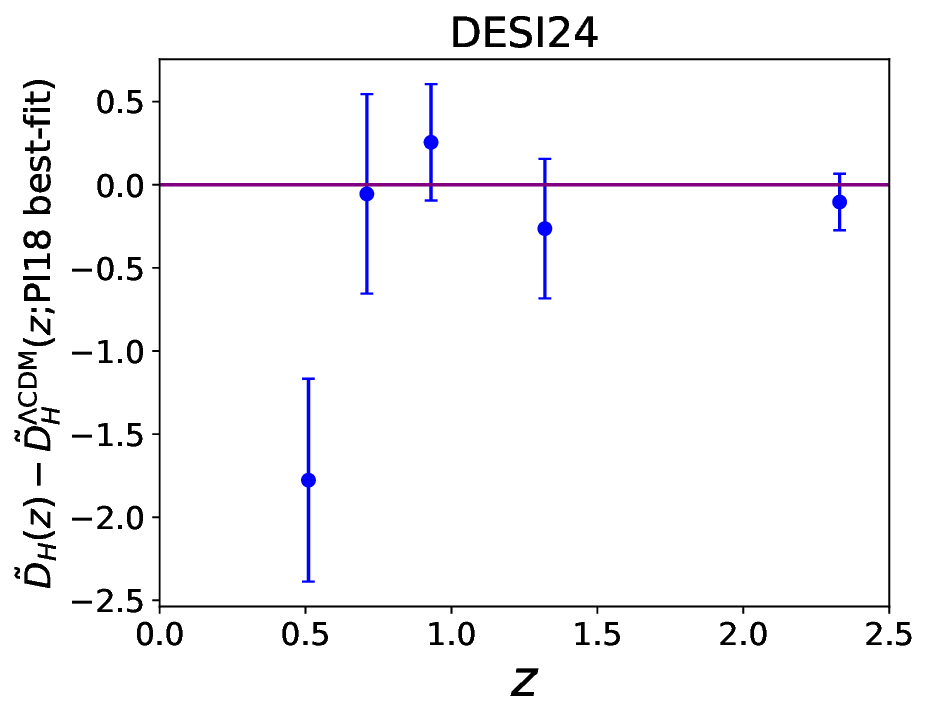}
\includegraphics[height=120pt,width=0.49\textwidth]{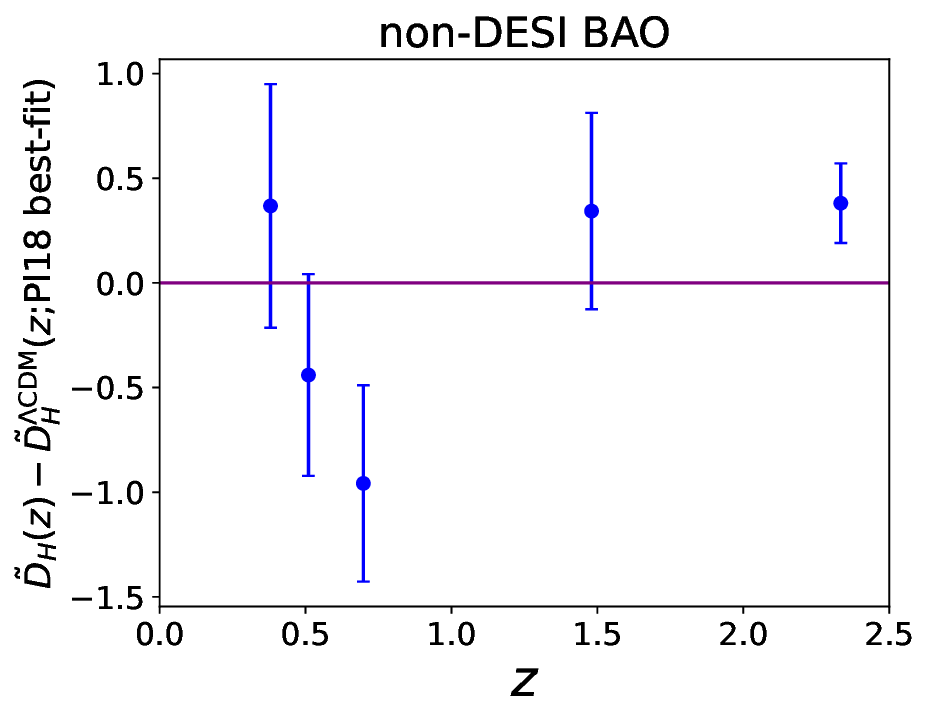} \\
\includegraphics[height=120pt,width=0.49\textwidth]{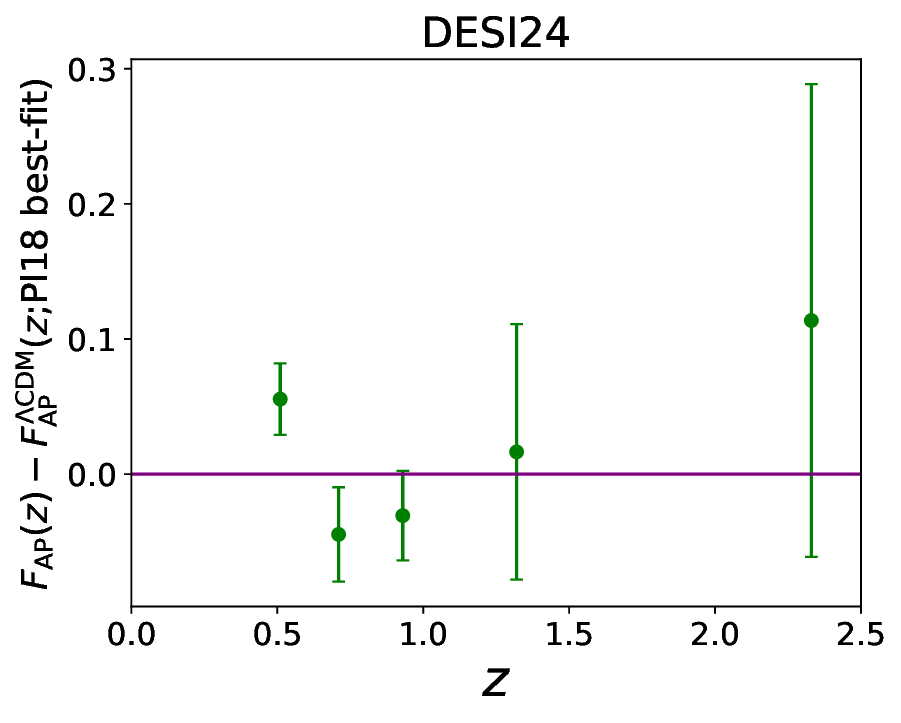}
\includegraphics[height=120pt,width=0.49\textwidth]{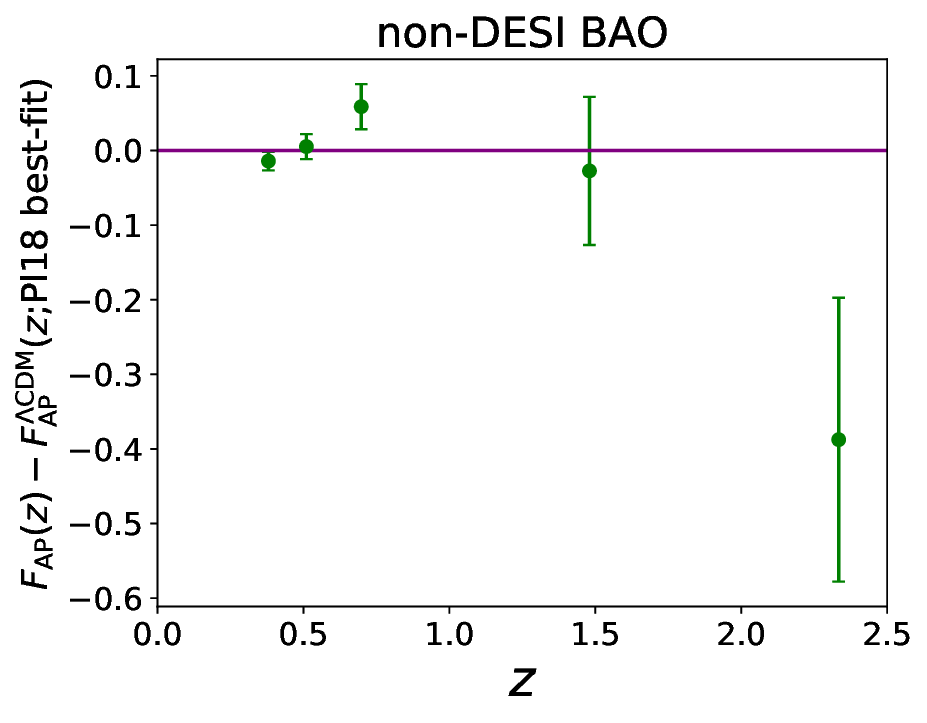} \\
\includegraphics[height=120pt,width=0.49\textwidth]{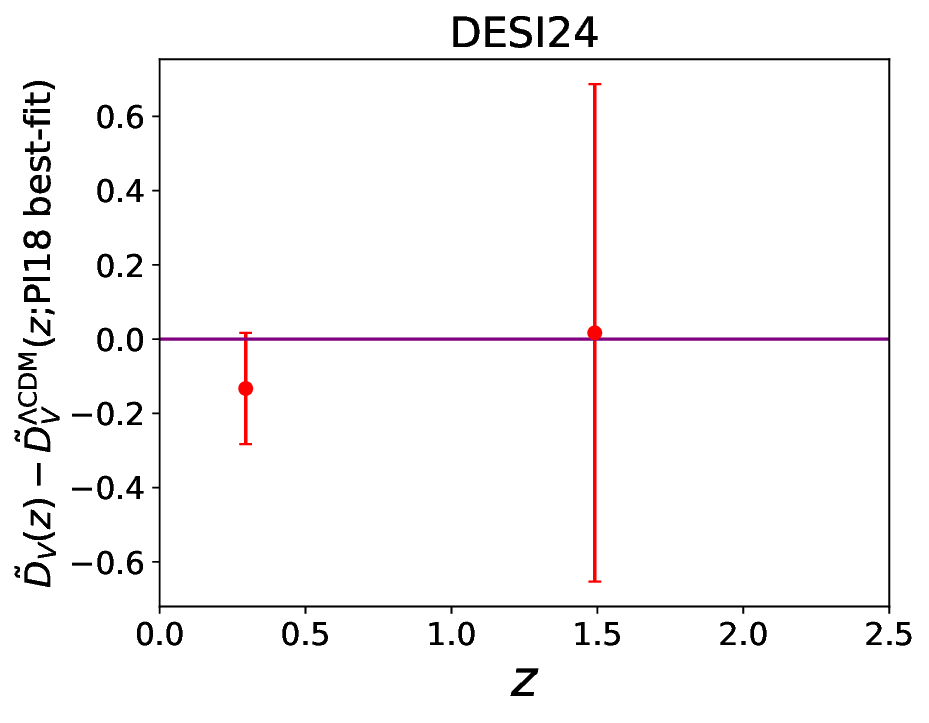}
\caption{
\label{fig:A_DESI_other_Lcdm}
Plots of $\tilde{D}_M-\tilde{D}_M^{\rm \Lambda CDM}$(Pl18 best-fit) (black), $\tilde{D}_H-\tilde{D}_H^{\rm \Lambda CDM}$(Pl18 best-fit) (blue), $F_{\rm AP}-F_{\rm AP}^{\rm \Lambda CDM}$(Pl18 best-fit) (green), and $\tilde{D}_V-\tilde{D}_V^{\rm \Lambda CDM}$(Pl18 best-fit) (red) both for DESI 2024 (left) and other BAO (right) data. Any deviation from the continuous horizontal lines (purple) i.e. zero values correspond to the deviation from the $\Lambda$CDM model with parameter values according to the Planck 2018 $\Lambda$CDM best-fit for TT,TE,EE+lowE+lensing (denoted as 'Pl18').
}
\end{figure}

In Table~\ref{table:DESI_24_FAP_data}, we list five different values of $F_{\rm AP}$ and the associated 1$\sigma$ error bars at five different effective redshifts obtained from the DESI 2024 data corresponding to anisotropic BAO observations \citep{DESI:2024mwx}, obtained using Eq.~\eqref{eq:defn_FAP}. The errors in $F_{\rm AP}$ are computed by the propagation of uncertainty given as

\begin{align}
\label{eq:delta_FAP}
\Delta F_{\rm AP} &= \sqrt{{\rm Var}[F_{\rm AP}]} , \nonumber\\
{\rm Var}[F_{\rm AP}] &= \left( \frac{\partial F_{\rm AP}}{\partial \tilde{D}_M} \right)^2 {\rm Var}[\tilde{D}_M] + \left( \frac{\partial F_{\rm AP}}{\partial \tilde{D}_H} \right)^2 {\rm Var}[\tilde{D}_H] + 2 \frac{\partial F_{\rm AP}}{\partial \tilde{D}_M} \frac{\partial F_{\rm AP}}{\partial \tilde{D}_H} r_{\rm MH} \Delta \tilde{D}_M \Delta \tilde{D}_H , \nonumber\\
\frac{\partial F_{\rm AP}}{\partial \tilde{D}_M} &= \frac{1}{\tilde{D}_H} , \nonumber\\
\frac{\partial F_{\rm AP}}{\partial \tilde{D}_H} &= - \frac{\tilde{D}_M}{\tilde{D}_H^2} ,
\end{align}

\noindent
where ${\rm Var}[F_{\rm AP}]$ means the variance of $F_{\rm AP}$. Note that, for simplicity to show the equations, we have omitted the redshift dependence in Eqs.~\eqref{eq:defn_rMH} and~\eqref{eq:delta_FAP}. It is understood that the redshift dependence is present in the relevant quantities. Using the above equations and using the values of $\tilde{D}_M$, $\tilde{D}_H$ and their correlations from Table~\ref{table:DESI24_DM_DH_tilde}, we compute $F_{\rm AP}$ and $\Delta F_{\rm AP}$ which are listed in Table~\ref{table:DESI_24_FAP_data}.

Throughout this analysis, we compute errors in any quantity through the (Gaussian) propagation of uncertainty with similar rules as in Eq.~\eqref{eq:delta_FAP}, where we take care of all the relevant self and cross covariances.

\subsubsection{Other BAO data}
\label{sec-other_BAO}

Besides, DESI 2024 data, we also consider other (non-DESI) BAO observations \citep{deCruzPerez:2024shj}. For this data, we closely follow Alam et al. (2020) \citep{eBOSS:2020yzd} which corresponds to the completed Sloan Digital Sky Survey (SDSS) IV. We consider five pairs of measurements of $\tilde{D}_M$ and $\tilde{D}_H$ at five effective redshifts. The first and second pairs correspond to the Baryon Oscillation Spectroscopic Survey (BOSS) DR12 LRG sample at effective redshifts 0.38 and 0.51 respectively \citep{BOSS:2016wmc}. The third pair corresponds to the extended Baryon Oscillation Spectroscopic Survey (eBOSS) DR16 LRG sample at effective redshift 0.698 \citep{eBOSS:2020yzd}. The fourth pair corresponds to the eBOSS DR16 QSO sample at effective redshift 1.48 \citep{eBOSS:2020gbb}. The fifth pair corresponds to eBOSS DR16 Ly-$\alpha$ QSO sample at effective redshift 2.334 \citep{eBOSS:2020tmo}. The values of $\tilde{D}_M$, $\tilde{D}_H$ and their correlations, obtained from other non-DESI BAO data are listed in Table~\ref{table:Other_BAO_DM_DH_tilde}. We denote these BAO data as 'non-DESI BAO'.

Using the same procedure as in the case of the DESI 2024 data, we compute $F_{\rm AP}$ and $\Delta F_{\rm AP}$ for non-DESI BAO observations which are listed in Table~\ref{table:Other_BAO_FAP_data}.

To check how close the values of $\tilde{D}_M$ (black), $\tilde{D}_H$ (blue), $F_{\rm AP}$ (green), and $\tilde{D}_V$ (red) compared to the reference $\Lambda$CDM model, we plot the differences in Fig.~\ref{fig:A_DESI_other_Lcdm} both for DESI 2024 (left panel) and other non-DESI BAO (right panel) data. For the reference model, we have considered the Planck 2018 best-fit $\Lambda$CDM for TT,TE,EE+lowE+lensing ($\Omega_{\rm m0}=0.3153$, $r_d=147.09$ Mpc, and $h=0.6736$). From the difference in $\tilde{D}_M$ for DESI 2024 data, we see that the reference $\Lambda$CDM model is slightly more than 2$\sigma$ away at effective redshift 0.71 and at other effective redshifts it is within 1$\sigma$. From the difference in $\tilde{D}_H$ for DESI 2024 data, we see that the reference $\Lambda$CDM model is slightly less than 3$\sigma$ away at effective redshift 0.51 and at other effective redshifts it is within 1$\sigma$. At all effective redshifts, it is within 1$\sigma$ for the differences in $F_{\rm AP}$ and $\tilde{D}_V$ for DESI24 data. For other non-DESI BAO data, reference $\Lambda$CDM model is within 1$\sigma$ (or at around 1$\sigma$) errors at most of the effective redshift points.

\subsection{Planck CMB mission 2018: CMB distance priors}
\label{sec-cmb_dist_prior}

The cosmic microwave background (CMB) observations can be summarised through two main quantities: the CMB shift parameter $R$ and the acoustic length scale $l_A$ at photon decoupling redshift $z_*$ \citep{Chen:2018dbv,Zhai:2019nad,Zhai:2018vmm}. The expressions of $R$ and $l_A$ are given as

\begin{eqnarray}
R &=& \frac{\sqrt{\Omega_{\rm m0}H_0^2}D_M(z_*)}{c},
\label{eq:CMB_R} \\
l_A &=& \frac{\pi D_M(z_*)}{r_s(z_*)},
\label{eq:CMB_lA}
\end{eqnarray}

\noindent
respectively, where $r_s$ is the sound horizon. For a late-time model-independent analysis, it is difficult to use these two quantities separately, because of the presence of the quantity $D_M$. In general, we can not get values of $D_M$ without knowing the form of $D_M$ corresponding to a particular model. However, the ratio of these two quantities is useful for a late-time model-independent analysis. This is the aim of this analysis. Dividing Eq.~\eqref{eq:CMB_R} by Eq.~\eqref{eq:CMB_lA}, we get the the ratio $Q$ given as

\begin{equation}
Q = \frac{R}{l_A} = \frac{\sqrt{\Omega_{\rm m0}H_0^2}r_s(z_*)}{c\pi} \approx \frac{\sqrt{\omega_{\rm m0}}}{20\pi} \frac{r_s(z_*)}{150~\rm Mpc} .
\label{eq:Q_R_by_lA}
\end{equation}

We include CMB data because our main requirement in this analysis is to get values of the $\omega_{\rm m0}$ parameter (for all four diagnostics) and the $r_d$ parameter (for diagnostics $A_1$, $A_2$ and $A_3$). If we consider standard early-time physics, both $r_s(z_*)$ and $r_d$ depend only on two parameters $\omega_{\rm m0}$ and $\omega_{\rm b0}$, where $\omega_{\rm b0}$ is given as

\begin{equation}
\omega_{\rm b0} = \Omega_{\rm b0} h^2 ,
\label{eq:defn_wb0}
\end{equation}

\noindent
with $\Omega_{\rm b0}$ being the baryon energy density parameter at present. For details, see Appendix~\ref{sec-appx_sound_horzon_early_standard}, Appendix~\ref{sec-appx_photon_decoupling_z_standard}, and Appendix~\ref{sec-appx_baryon_drag_z_standard}. Consequently, through Eq.~\eqref{eq:Q_R_by_lA}, for the standard early-time physics, $Q$ also depends only on these two parameters i.e.

\begin{equation}
Q \equiv Q(\omega_{\rm m0},\omega_{\rm b0}) .
\label{eq:Q_R_by_lA_2}
\end{equation}

\noindent
So, to get values of $\omega_{\rm m0}$ and $r_d$, we need another equation which is determined by fixing $\omega_{\rm b0}$ parameter. We consider values of $Q$ and $\omega_{\rm b0}$, their errors, and their correlation coefficient according to Planck 2018 results for TT,TE,EE+lowE+lensing. These are mentioned in Appendix~\ref{sec-cmb_dist_prior_again}. We call this CMB data the 'Pl18(standard)' throughout this analysis.

One can also go beyond the standard early-time physics approximation. For this case, for each extra degree of freedom, we need each extra equation for CMB data. Note that, for this case, the $Q$ parameter depends on these extra degrees of freedom through extra parameters i.e. Eq.~\eqref{eq:Q_R_by_lA_2} gets modified accordingly for non-standard early physics. We consider a non-standard early cosmology with two extra degrees of freedom by allowing variations in the sum of neutrino masses $\Sigma m_{\nu}$ and effective number of relativistic species $N_{\rm eff}$. So, we need two extra equations and for this, we follow Zhai et al. (2020) \citep{Zhai:2019nad}. These are mentioned in Appendix~\ref{sec-cmb_dist_prior_again} with the same CMB data. We call this CMB data as 'Pl18(standard+$\Sigma m_{\nu}$+$N_{\rm eff}$)' throughout the analysis.

\begin{table}
\begin{center}
\begin{tabular}{|c|c|c|c|c|}
\hline
Observation & $\omega_{\rm m0} \pm \Delta \omega_{\rm m0}$ & $r_d \pm \Delta r_d$ [Mpc] & $r(\omega_{\rm m0},r_d/{\rm Mpc})$ \\
\hline
Pl18(standard) & $0.1430 \pm 0.0011$ & $146.995 \pm 0.264$ & $-0.90$ \\
Pl18(standard+$\Sigma m_{\nu}$+$N_{\rm eff}$) & $0.1411 \pm 0.0041$ & $147.70 \pm 2.50$ & $-0.97$ \\
\hline
\end{tabular}
\end{center}
\caption{
The values of $\omega_{\rm m0}$ and $r_d$ parameters, errors in these, and correlation coefficient between these corresponding to Planck 2018 results for TT,TE,EE+lowE+lensing both for standard early time physics and a non-standard early cosmology by allowing variations in sum of neutrino masses $\Sigma m_{\nu}$ and effective number of relativistic species $N_{\rm eff}$.
}
\label{table:wm0_rd}
\end{table}

Using all these equations, we find $\omega_{\rm m0}$ and $r_d$, their errors, and their correlation coefficient. We list these values in Table~\ref{table:wm0_rd} both for standard and non-standard early-time physics.

Note that the constraints mentioned in Table~\ref{table:wm0_rd} are mostly independent of late-time dark energy models but the dark energy dependence may arise if the expansion history at late time in a dark energy model significantly deviates from the standard $\Lambda$CDM model.

\subsection{$H_0$ measurements}
\label{sec-H0_data}

To compute the diagnostics, we need the values of the $h$ parameter unless we get $\Omega_{\rm m0}$ directly (which is difficult to get directly since this is a derived parameter). To the best of our knowledge, we believe, there are no such observations which directly provide $\Omega_{\rm m0}$. CMB observations (or any other alternative useful observations) provide the combination $\Omega_{\rm m0}h^2$, not directly $\Omega_{\rm m0}$. So, to break the degeneracy of $\Omega_{\rm m0}h^2$ to get $\Omega_{\rm m0}$, we require the value of $h$. Hence, we need the values of the $H_0$ parameter. For this purpose, we consider two observations: the tip of the Red Giant Branch (tRGB) which corresponds to $H_0=69.8\pm 1.9$ [Km s$^{-1}$ Mpc$^{-1}$] i.e. $h=0.698\pm 0.019$ \citep{Freedman:2019jwv} and the SHOES observations which correspond to $H_0=73.2\pm 1.3$ [Km s$^{-1}$ Mpc$^{-1}$] i.e. $h=0.732\pm 0.013$ \citep{Riess:2020fzl}. We call these data as '$H_0$(tRGB)' and '$H_0$(SHOES)' respectively.

\begin{table}
\begin{center}
\begin{tabular}{|c|c|c|}
\hline
Observation & $h \pm \Delta h$ & Refs. \\
\hline
$H_0$(tRGB) & $0.698\pm 0.019$ & \citep{Freedman:2019jwv} \\
$H_0$(SHOES) & $0.732\pm 0.013$ & \citep{Riess:2020fzl} \\
\hline
\end{tabular}
\end{center}
\caption{
The two values of $h$, obtained from tRGB and SHOES data, denoted by '$H_0$(tRGB)' and '$H_0$(SHOES)' respectively.
}
\label{table:h_from_H0_data}
\end{table}

The $h$ values are listed in Table~\ref{table:h_from_H0_data}.

\begin{figure*}
\centering
\includegraphics[height=130pt,width=0.49\textwidth]{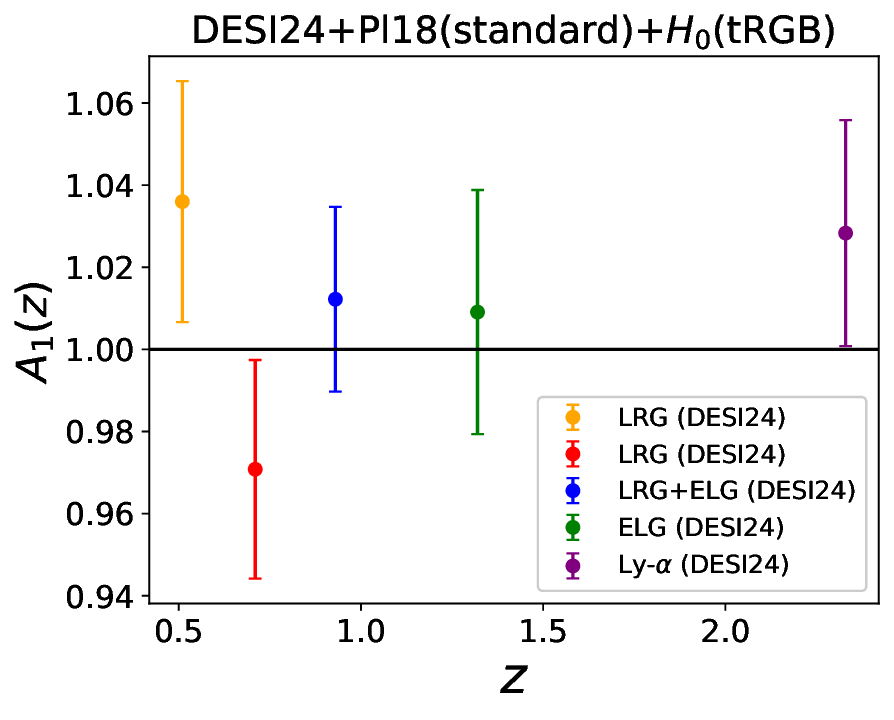}
\includegraphics[height=130pt,width=0.49\textwidth]{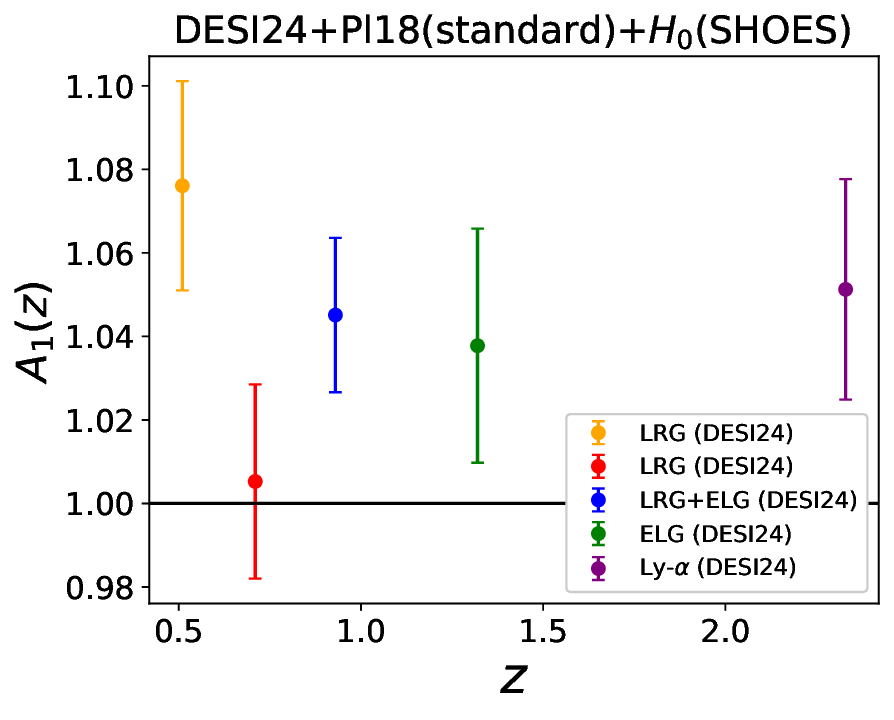} \\
\includegraphics[height=130pt,width=0.49\textwidth]{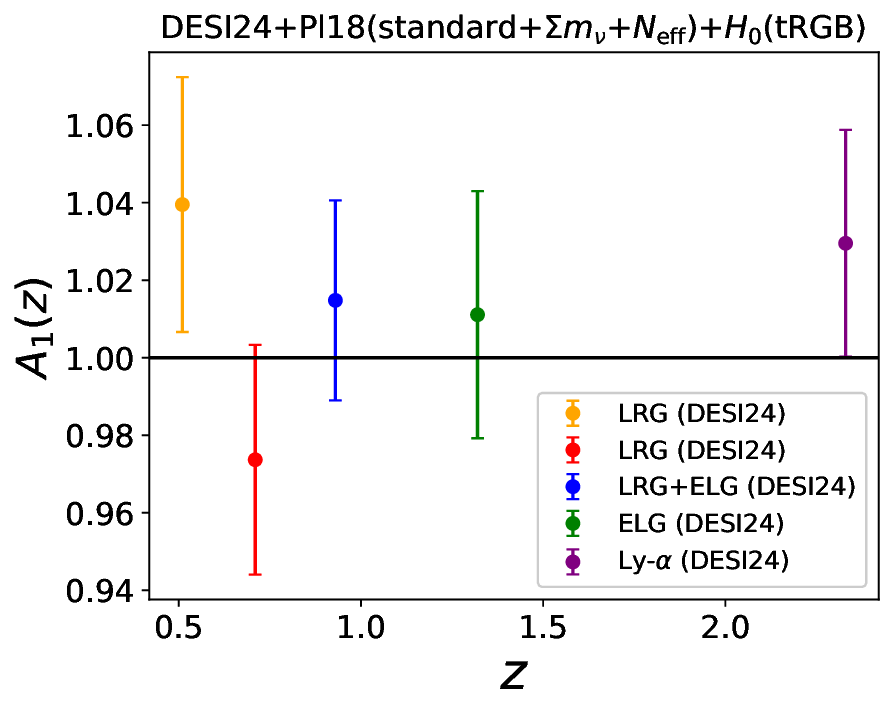}
\includegraphics[height=130pt,width=0.49\textwidth]{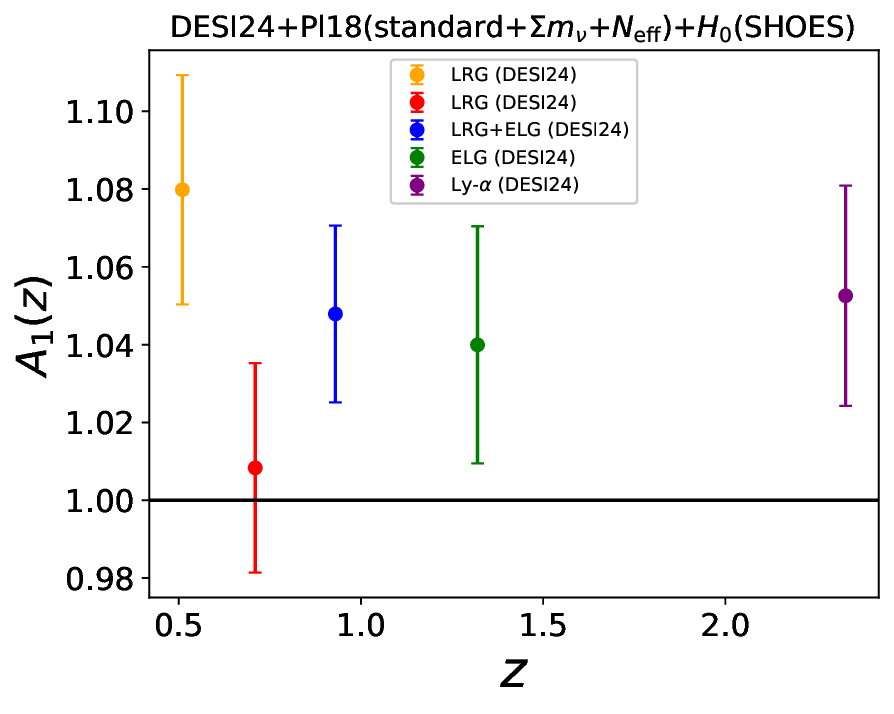}
\caption{
\label{fig:DESI_A1}
Plots of the diagnostic variable $A_1$ and the associated 1$\sigma$ errors. These values are obtained from DESI 2024 (DESI24) data combined with CMB and $H_0$ measurements using Eq.~\eqref{eq:diagnostic_A1} (with $\tilde{D}_M$ values obtained from Table~\ref{table:DESI24_DM_DH_tilde}, $h$ values obtained from Table~\ref{table:h_from_H0_data}, $\omega_{\rm m0}$ and $r_d$ values obtained from Table~\ref{table:wm0_rd}).
}
\end{figure*}

\begin{figure*}
\centering
\includegraphics[height=130pt,width=0.49\textwidth]{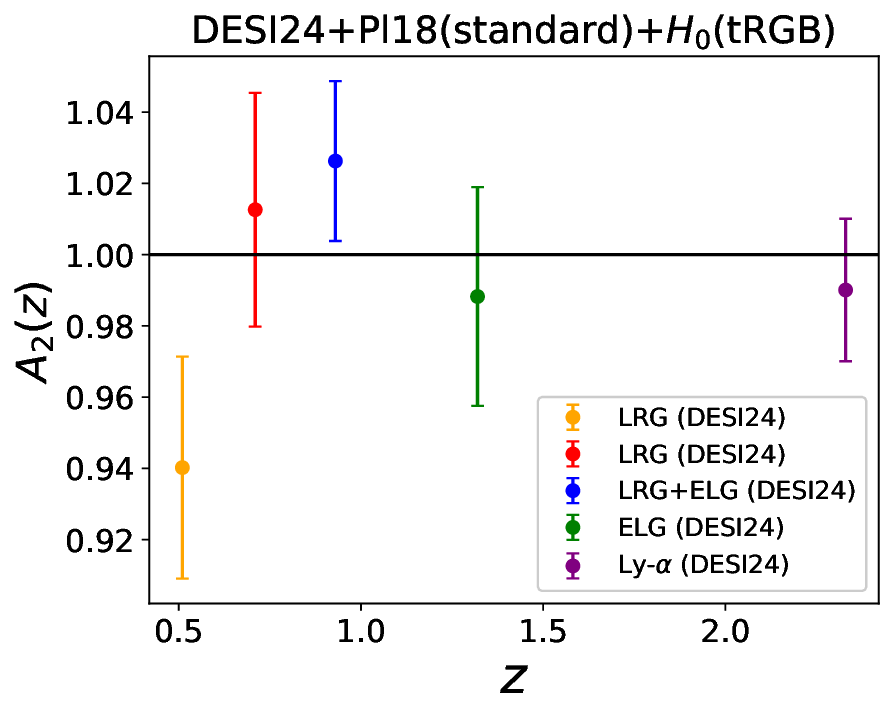}
\includegraphics[height=130pt,width=0.49\textwidth]{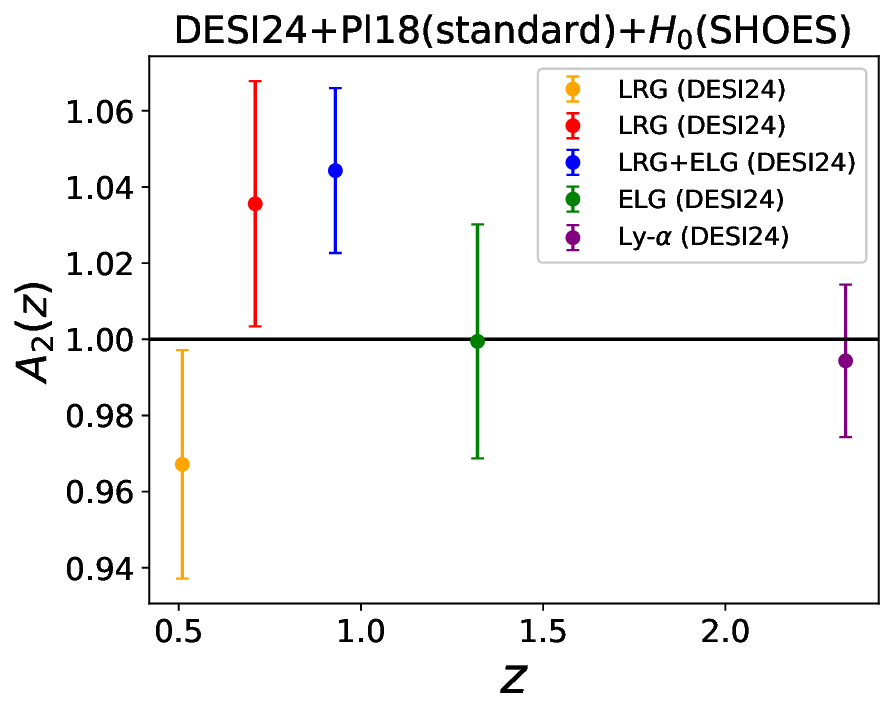} \\
\includegraphics[height=130pt,width=0.49\textwidth]{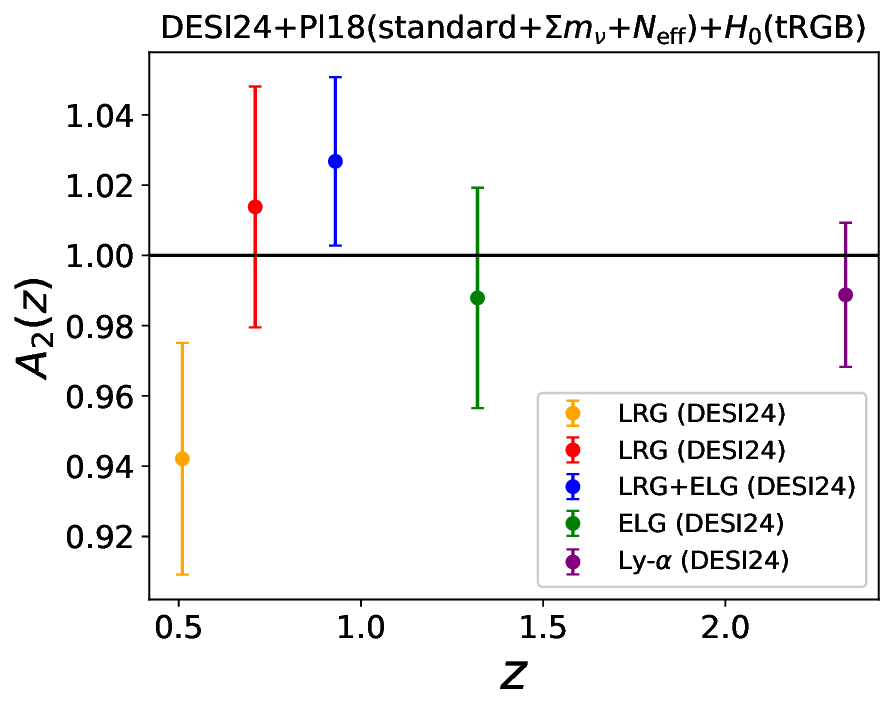}
\includegraphics[height=130pt,width=0.49\textwidth]{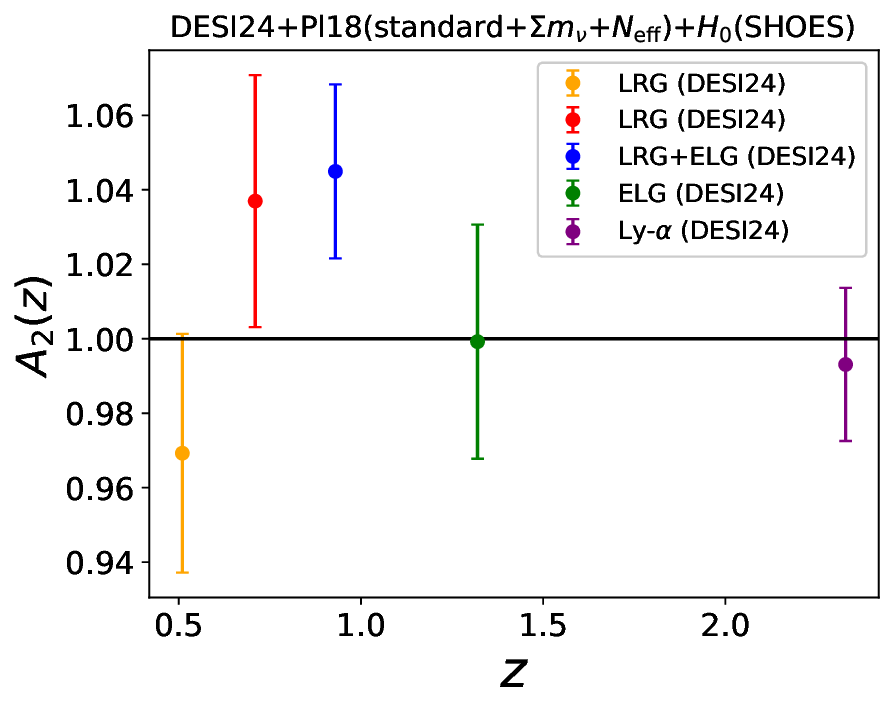}
\caption{
\label{fig:DESI_A2}
Diagnostic variable $A_2$ and the associated 1$\sigma$ errors for DESI 2024 BAO obtained using Eq.~\eqref{eq:diagnostic_A2} (with $\tilde{D}_H$ values obtained from Table~\ref{table:DESI24_DM_DH_tilde}).
}
\end{figure*}

\begin{figure*}
\centering
\includegraphics[height=120pt,width=0.49\textwidth]{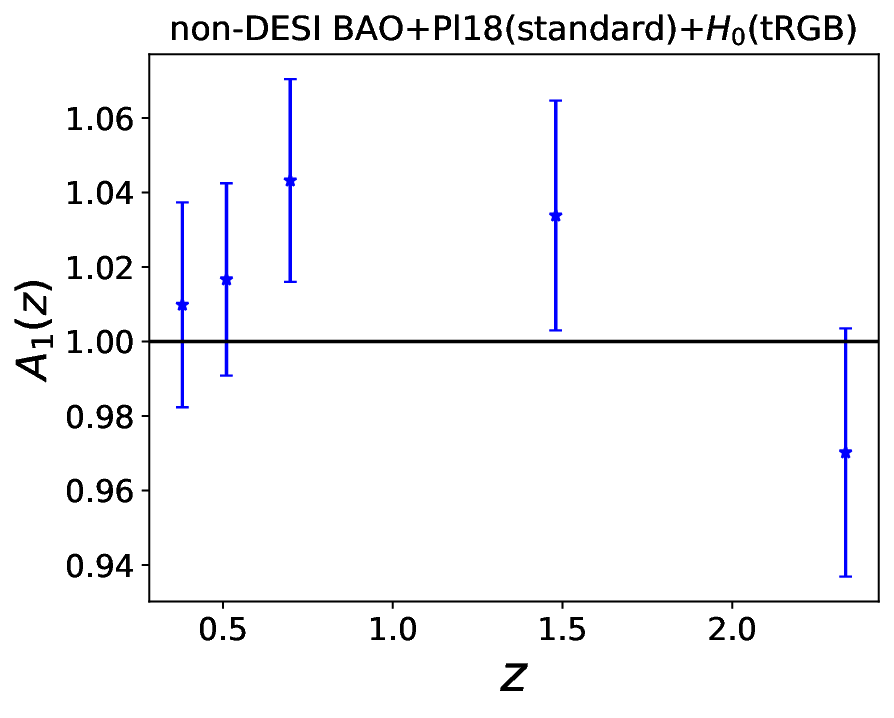}
\includegraphics[height=120pt,width=0.49\textwidth]{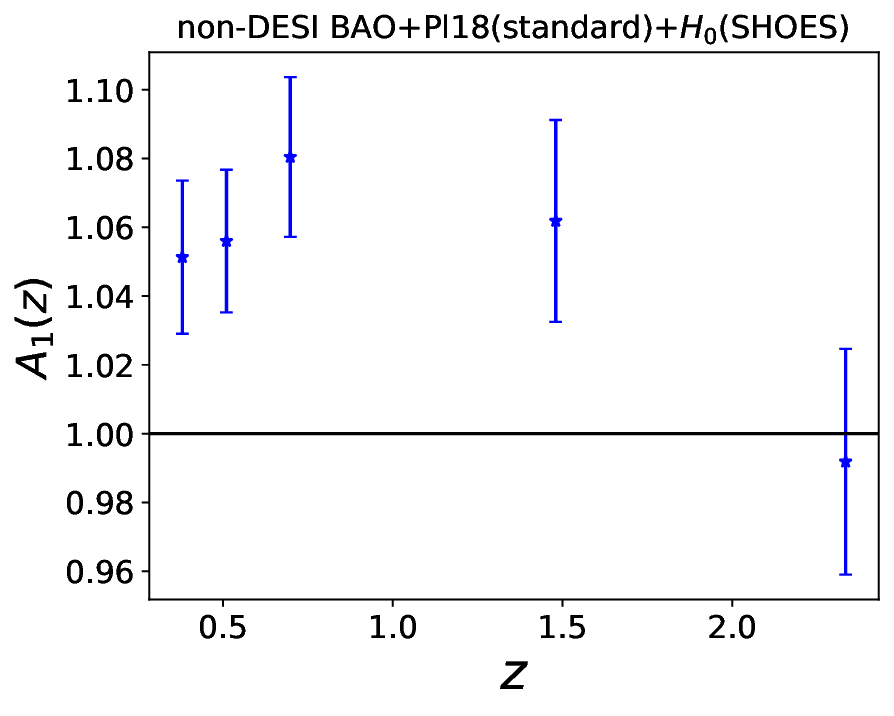} \\
\includegraphics[height=120pt,width=0.49\textwidth]{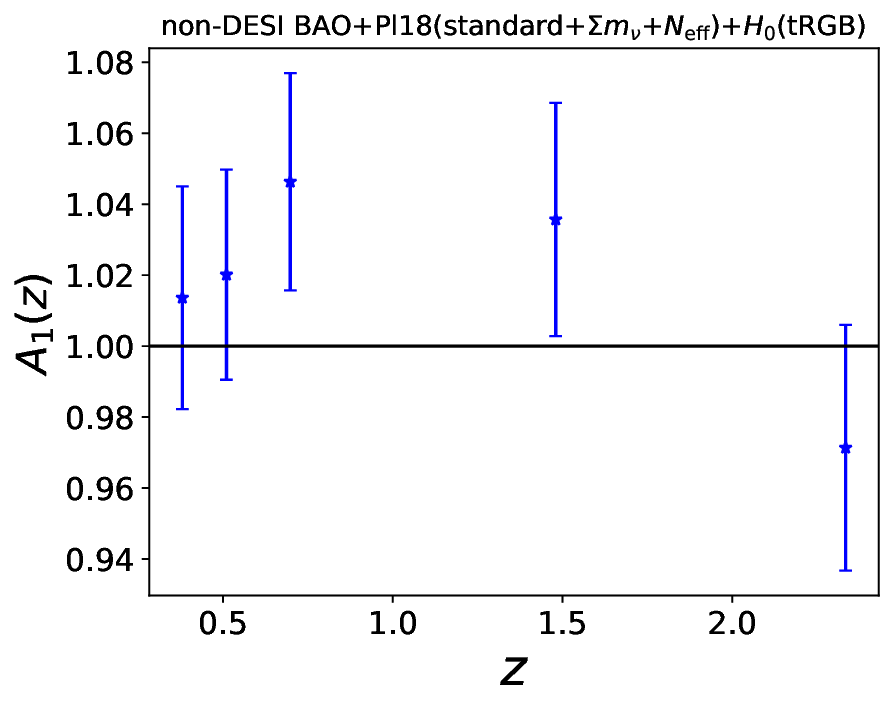}
\includegraphics[height=120pt,width=0.49\textwidth]{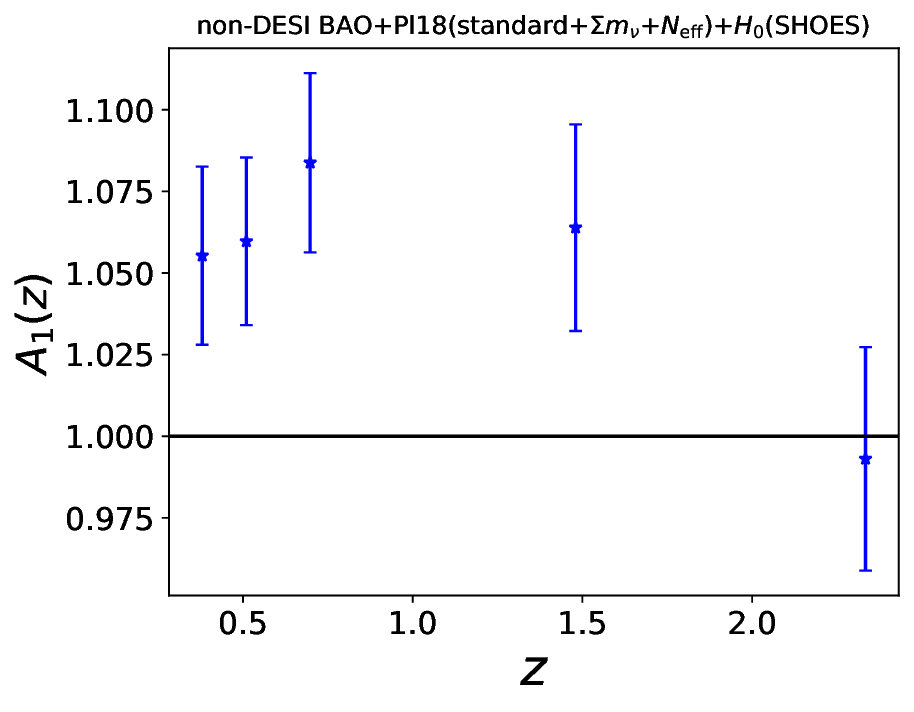}
\caption{
\label{fig:nondesi_A1}
The diagnostic variable $A_1$ and the associated 1$\sigma$ errors for non-DESI BAO data.
}
\end{figure*}

\begin{figure*}
\centering
\includegraphics[height=120pt,width=0.49\textwidth]{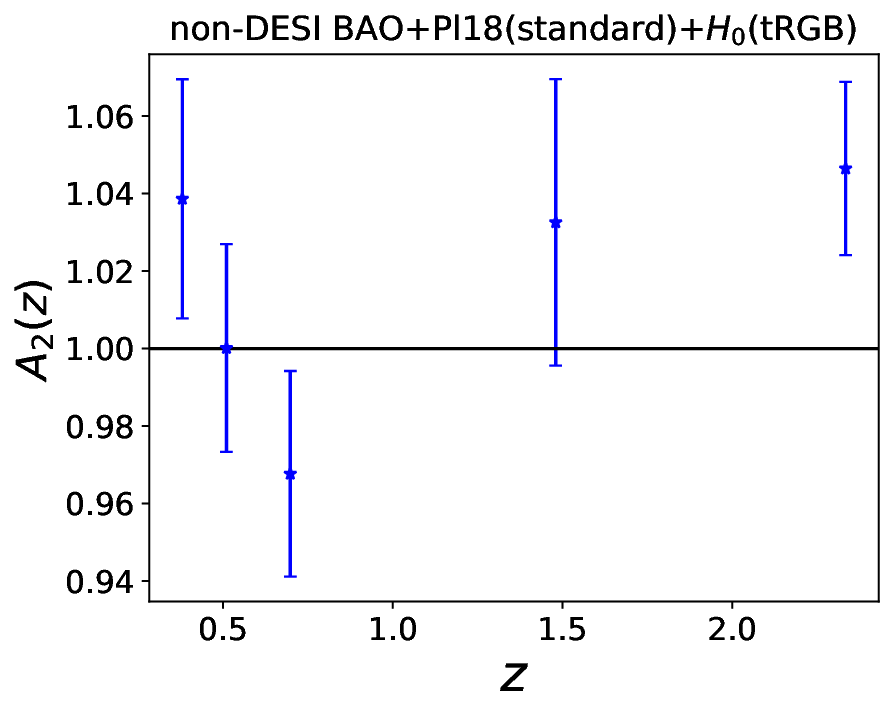}
\includegraphics[height=120pt,width=0.49\textwidth]{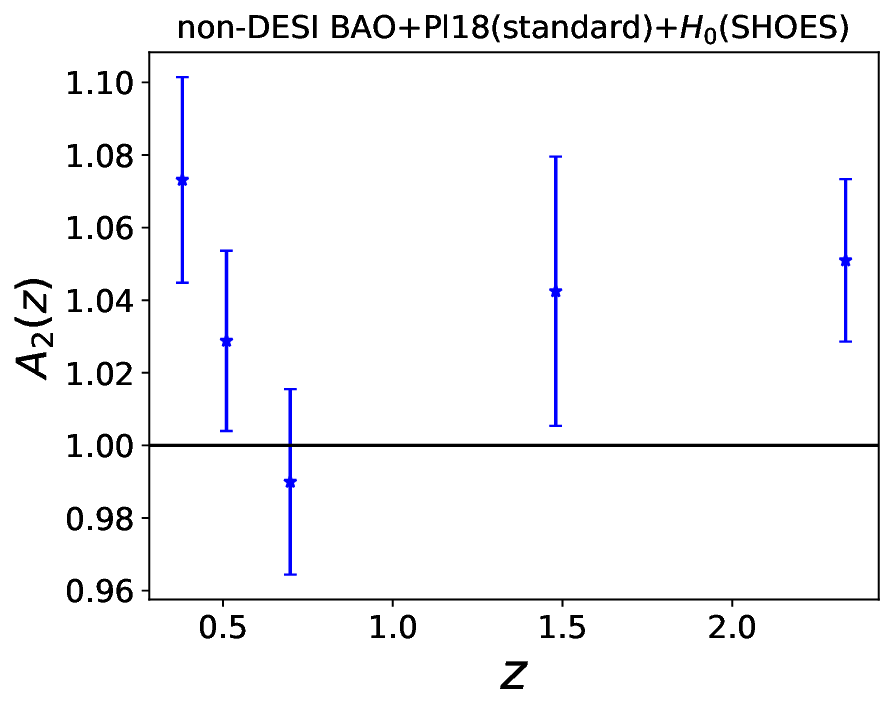} \\
\includegraphics[height=120pt,width=0.49\textwidth]{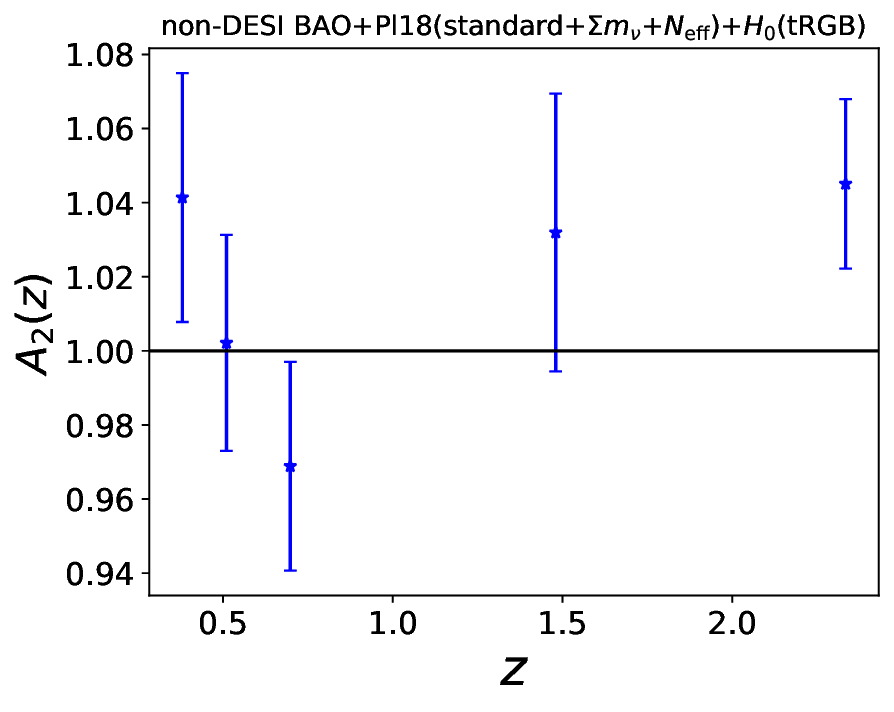}
\includegraphics[height=120pt,width=0.49\textwidth]{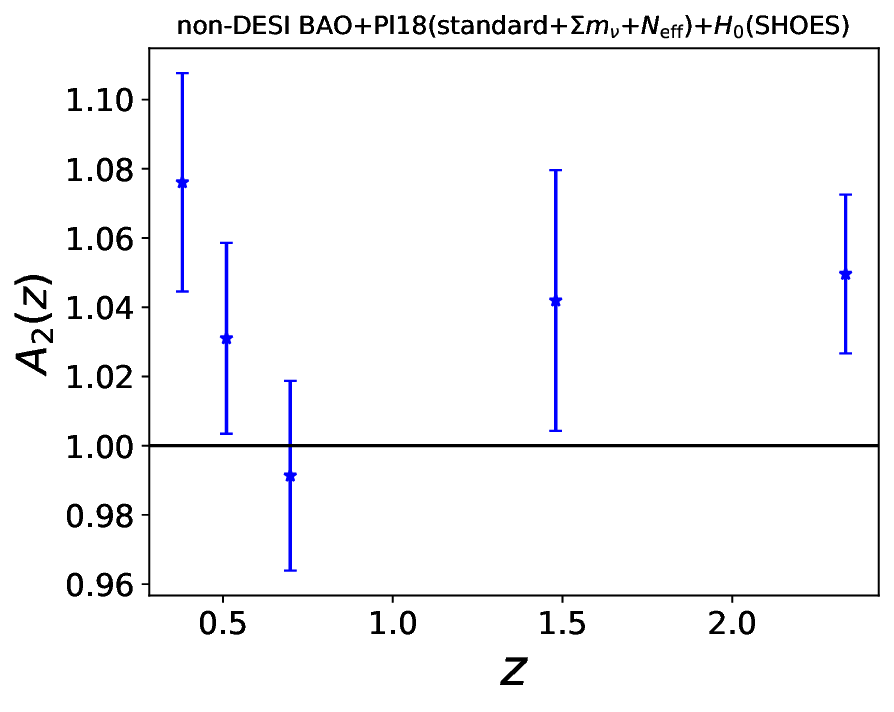}
\caption{
\label{fig:nondesi_A2}
The diagnostic variable $A_2$ and the associated 1$\sigma$ errors for non-DESI BAO data.
}
\end{figure*}

\begin{figure*}
\centering
\includegraphics[height=130pt,width=0.49\textwidth]{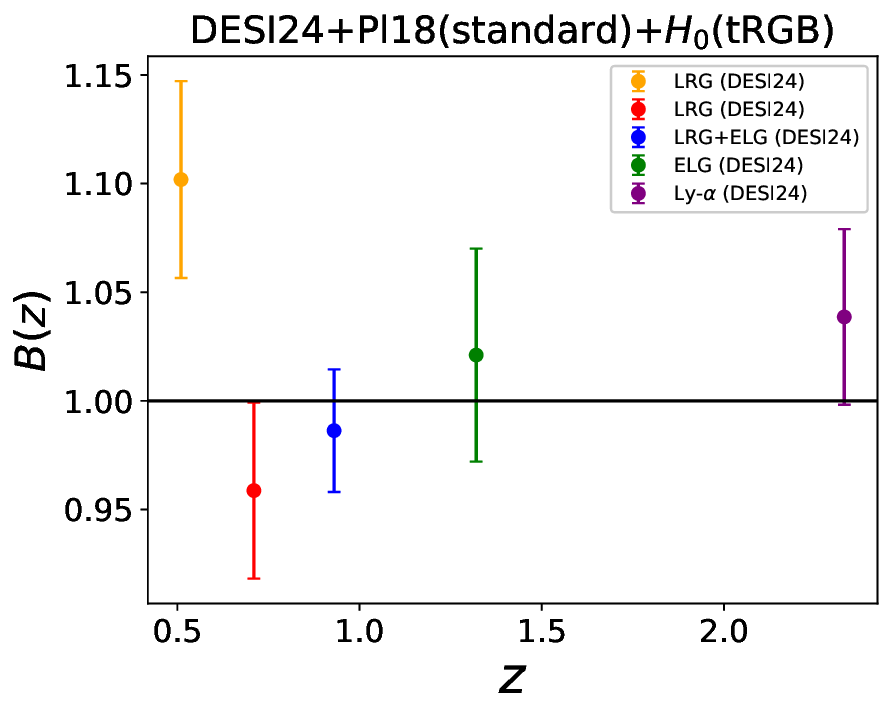}
\includegraphics[height=130pt,width=0.49\textwidth]{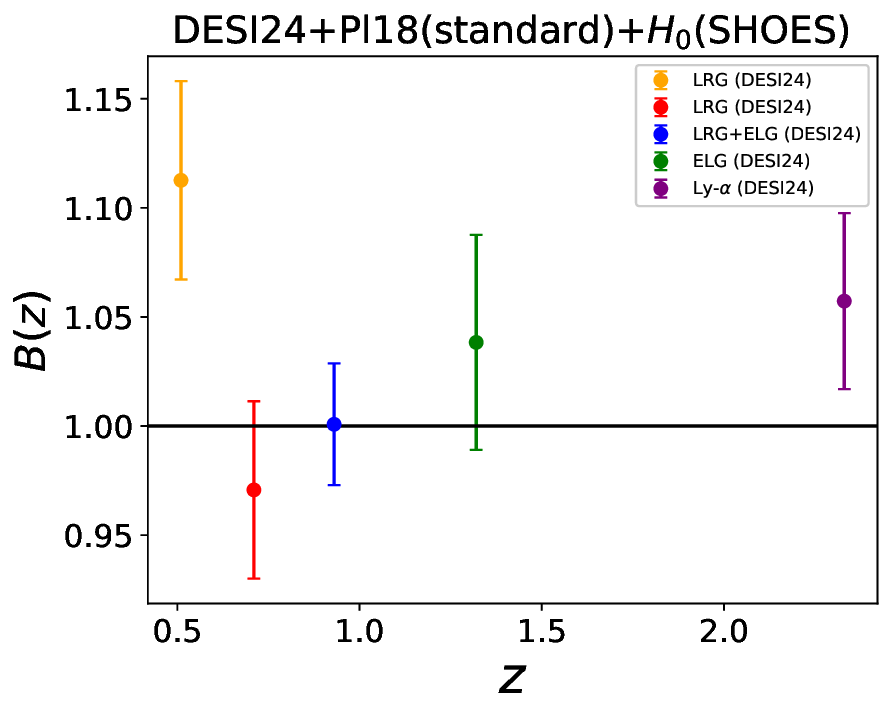} \\
\includegraphics[height=130pt,width=0.49\textwidth]{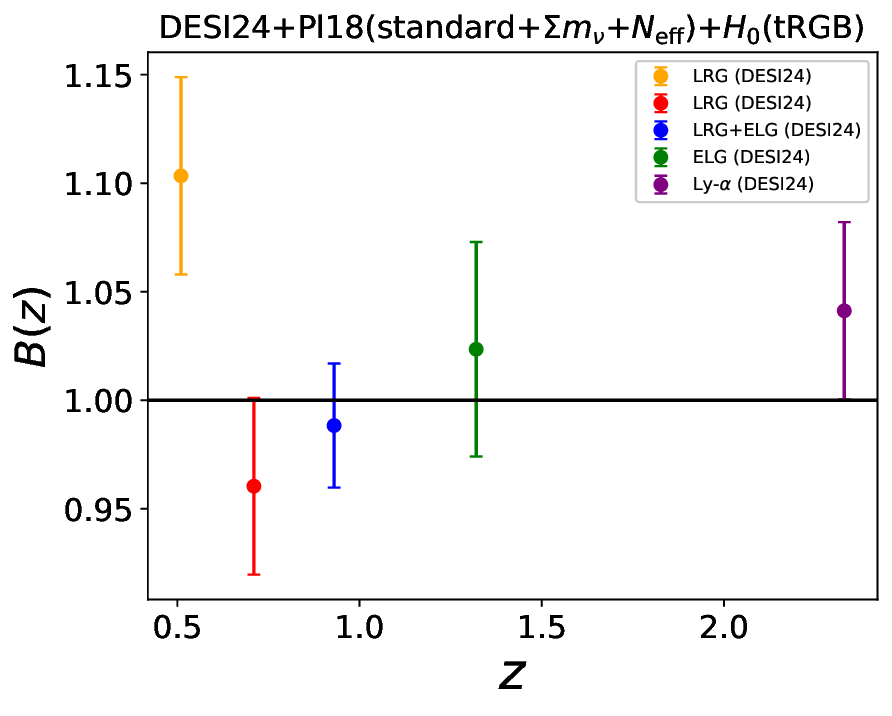}
\includegraphics[height=130pt,width=0.49\textwidth]{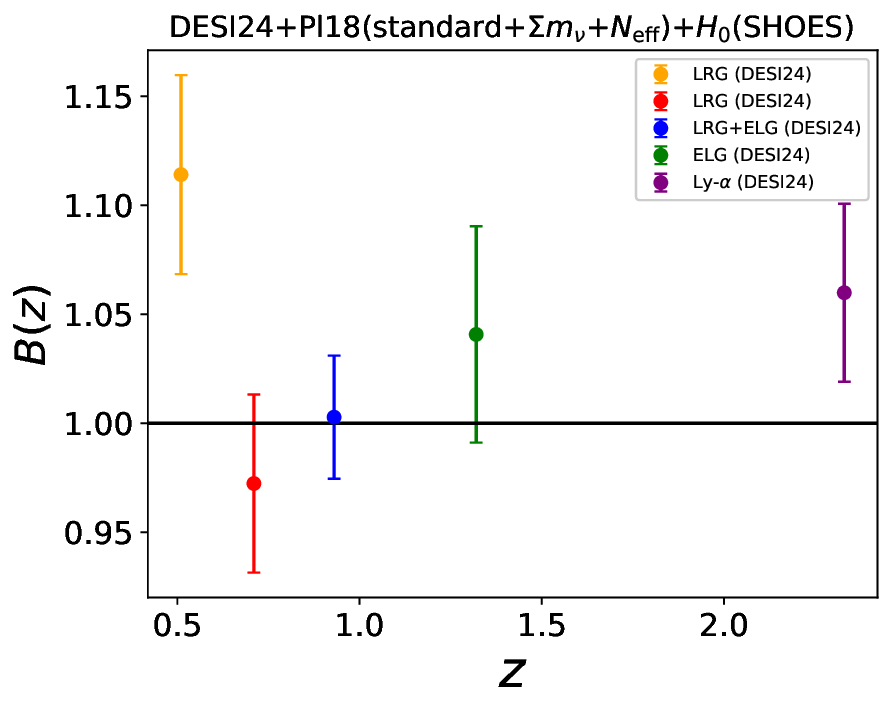}
\caption{
\label{fig:B_check_DESI}
Plots of the diagnostic variable $B$ and the associated 1$\sigma$ errors for DESI 2024 BAO data combined with CMB and $H_0$ measurements using Eq.~\eqref{eq:defn_B} (with $F_{\rm AP}$ values obtained from Table~\ref{table:DESI_24_FAP_data}).
}
\end{figure*}

\begin{figure*}
\centering
\includegraphics[height=120pt,width=0.49\textwidth]{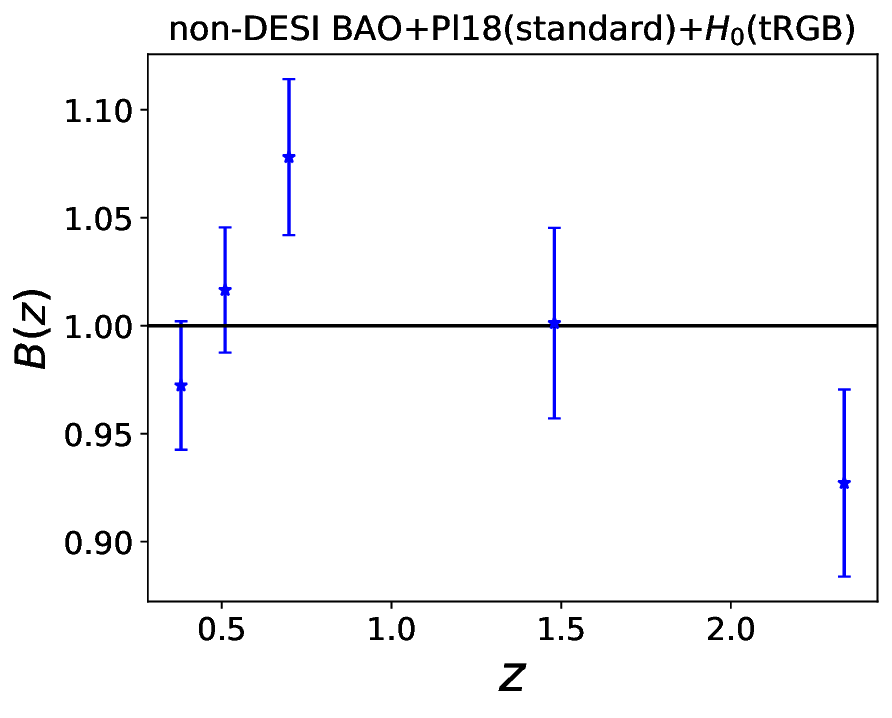}
\includegraphics[height=120pt,width=0.49\textwidth]{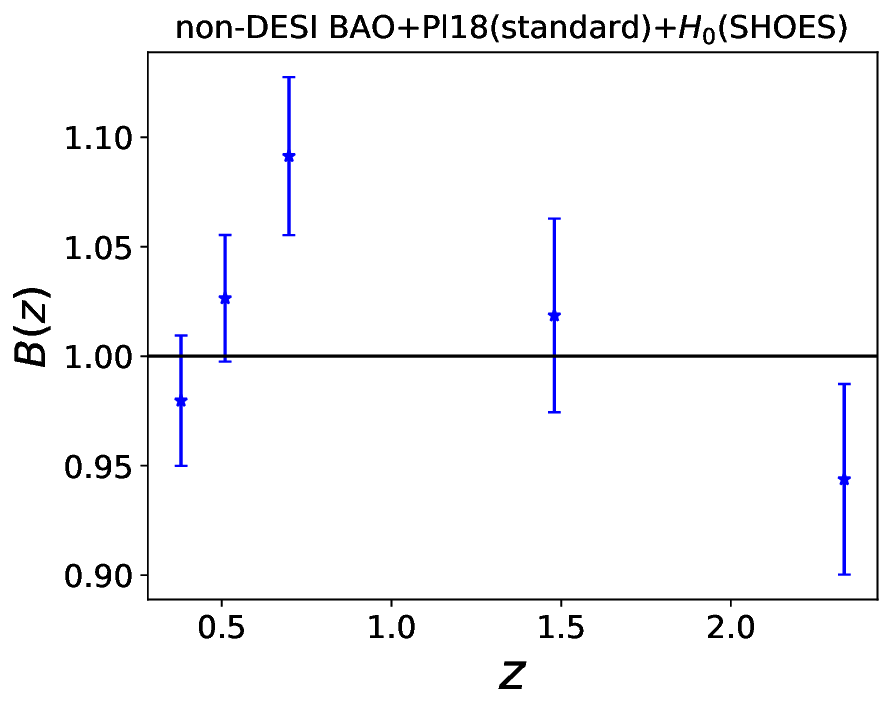} \\
\includegraphics[height=120pt,width=0.49\textwidth]{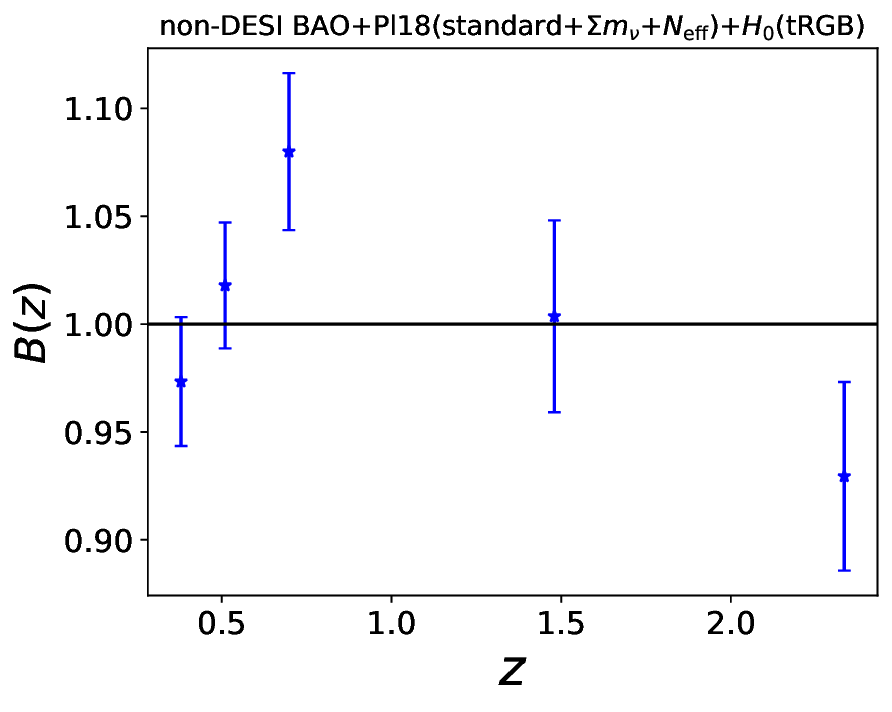}
\includegraphics[height=120pt,width=0.49\textwidth]{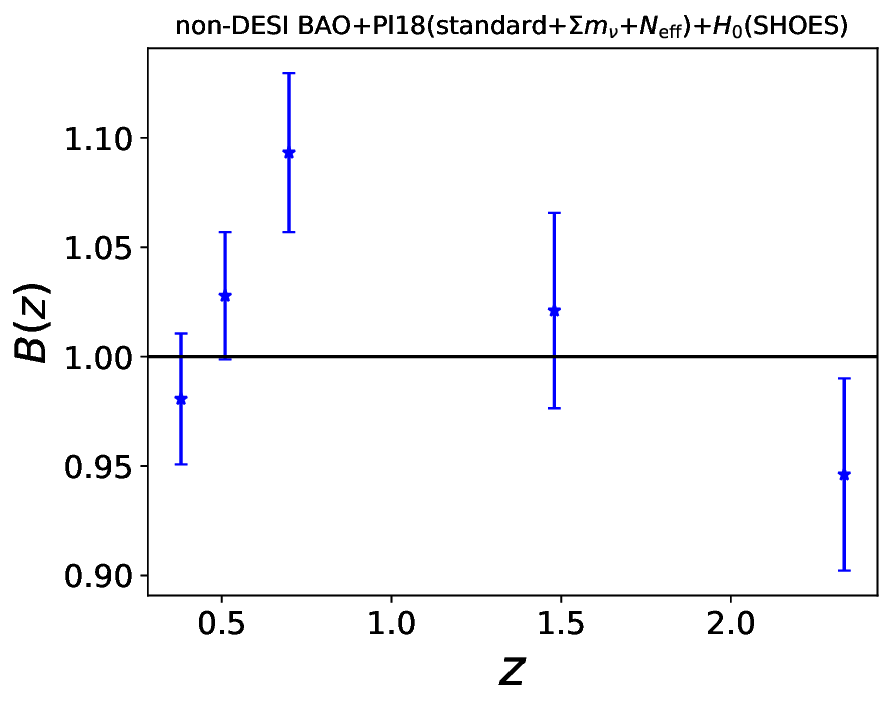}
\caption{
\label{fig:B_non_DESI_BAO}
The diagnostic variable $B$ and the associated 1$\sigma$ errors for non-DESI BAO data.
}
\end{figure*}

\begin{figure*}
\centering
\includegraphics[height=120pt,width=0.49\textwidth]{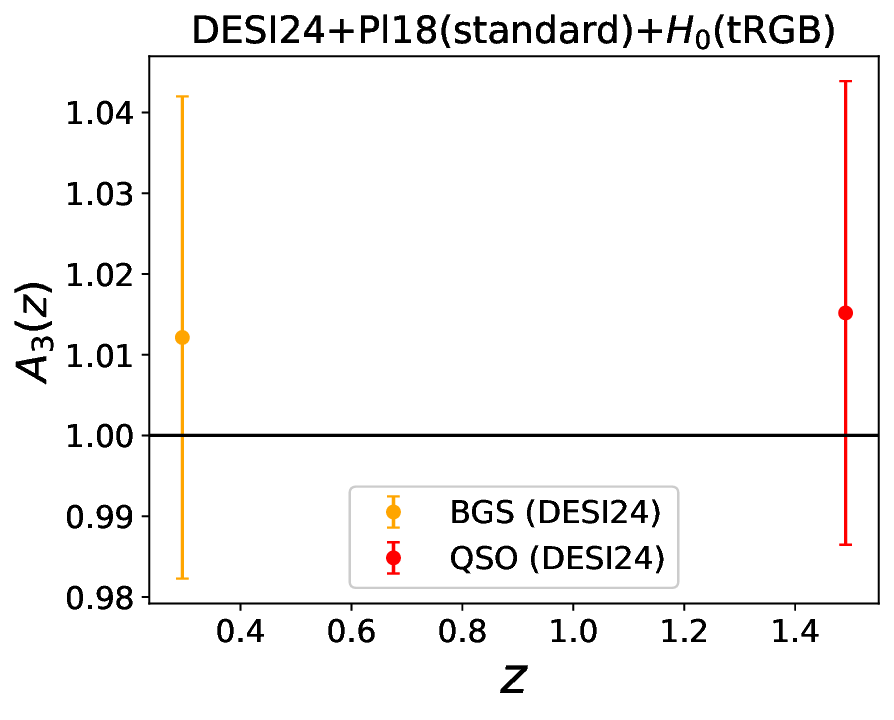}
\includegraphics[height=120pt,width=0.49\textwidth]{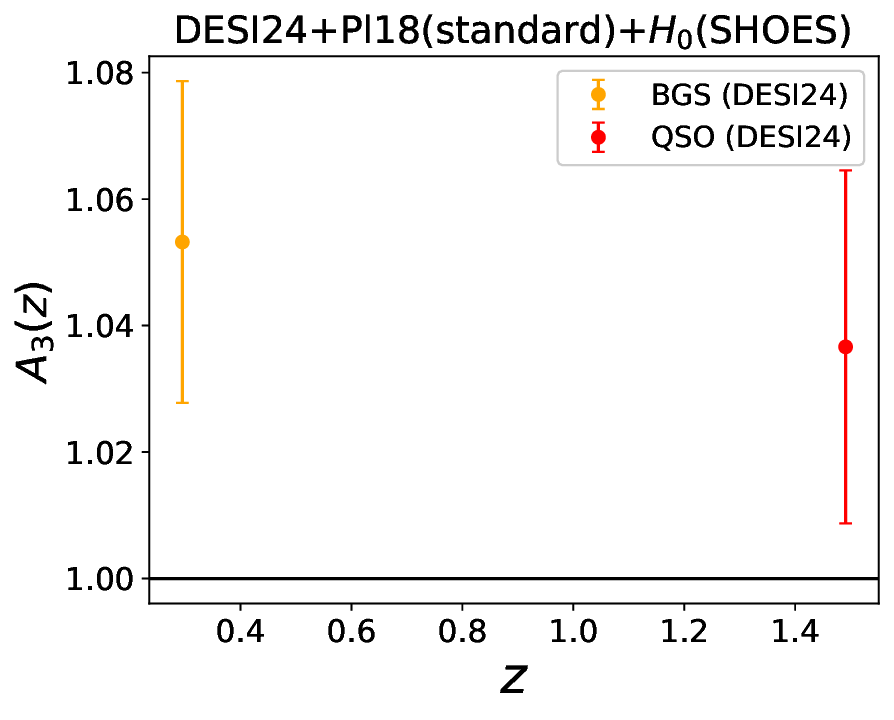} \\
\includegraphics[height=120pt,width=0.49\textwidth]{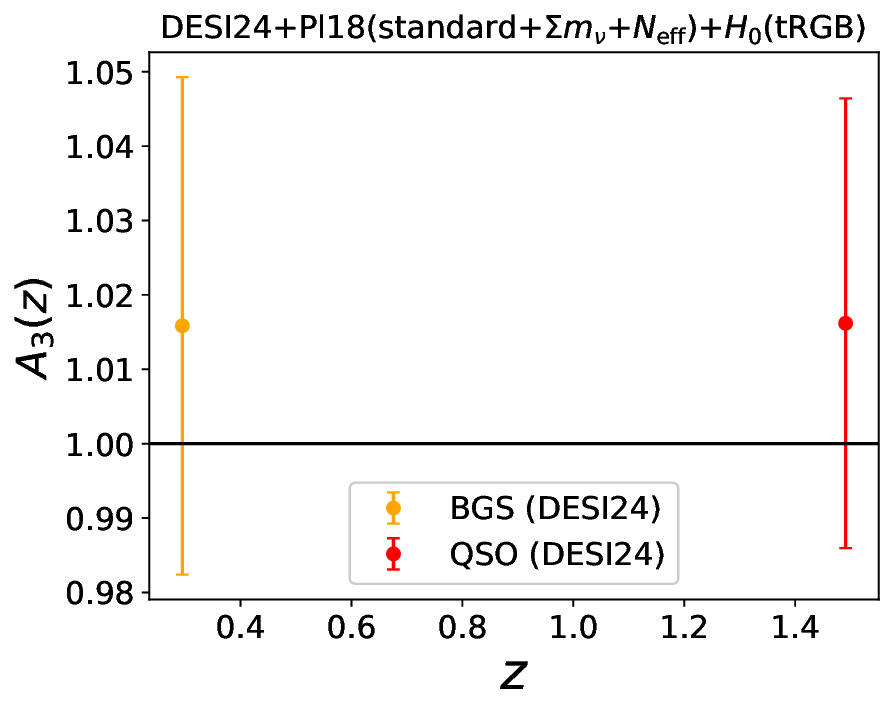}
\includegraphics[height=120pt,width=0.49\textwidth]{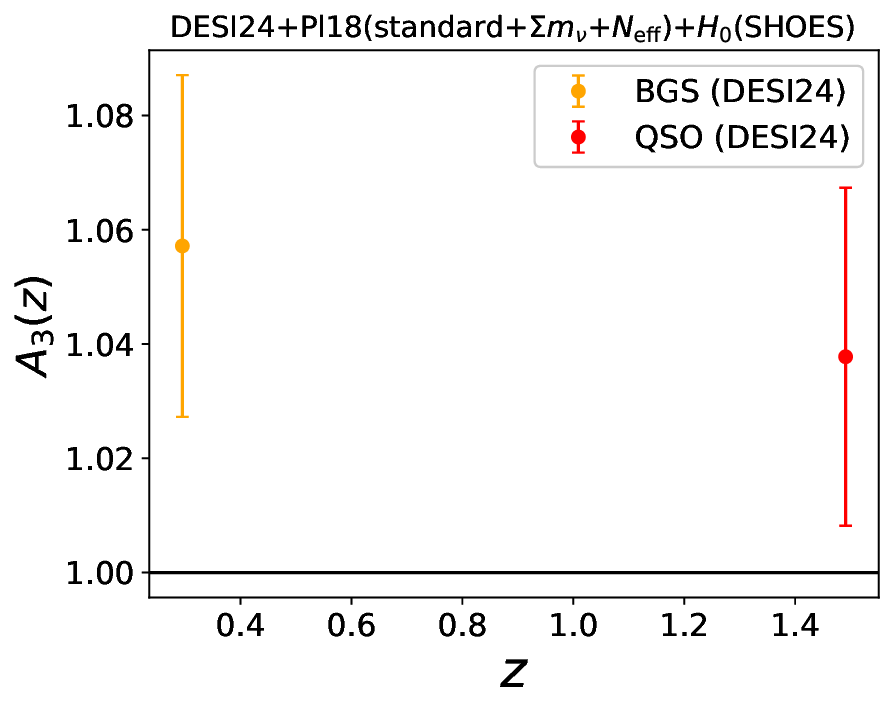}
\caption{
\label{fig:desi_plot_A3}
Plots of the diagnostic variable $A_3(z)$ and the associated 1$\sigma$ errors. These values are obtained from DESI 2024 data combined with CMB and $H_0$ measurements using Eq.~\eqref{eq:diagnostic_A3} (with $\tilde{D}_V$ values obtained from Table~\ref{table:DESI24_DM_DH_tilde}).
}
\end{figure*}

\begin{table*}
\begin{center}
DESI24+Pl18(standard) \\
\begin{tabular}{ |c|c|c|c|c| }
\hline
Data & $\chi^2$ ($A_1\&A_2$) & dof ($A_1\&A_2$) & p ($A_1\&A_2$) & CL ($A_1\&A_2$) \\
\hline
+$H_0$(tRGB) & 13.456 & 10 & 0.199 & 80.07\% \\
+$H_0$(SHOES) & 19.881 & 10 & 0.030 & 96.96\% \\
\hline
\end{tabular} \\
DESI24+Pl18(standard+$\Sigma m_{\nu}$+$N_{\rm eff}$) \\
\begin{tabular}{ |c|c|c|c|c| }
\hline
Data & $\chi^2$ ($A_1\&A_2$) & dof ($A_1\&A_2$) & p ($A_1\&A_2$) & CL ($A_1\&A_2$) \\
\hline
+$H_0$(tRGB) & 13.225 & 10 & 0.211 & 78.86\% \\
+$H_0$(SHOES) & 16.641 & 10 & 0.083 & 91.73\% \\
\hline
\end{tabular} \\
DESI24+Pl18(standard) \\
\begin{tabular}{ |c|c|c|c|c| }
\hline
Data & $\chi^2$ ($B$) & dof ($B$) & p ($B$) & CL ($B$) \\
\hline
+$H_0$(tRGB) & 7.565 & 5 & 0.182 & 81.81\% \\
+$H_0$(SHOES) & 9.113 & 5 & 0.105 & 89.53\% \\
\hline
\end{tabular} \\
DESI24+Pl18(standard+$\Sigma m_{\nu}$+$N_{\rm eff}$) \\
\begin{tabular}{ |c|c|c|c|c| }
\hline
Data & $\chi^2$ ($B$) & dof ($B$) & p ($B$) & CL ($B$) \\
\hline
+$H_0$(tRGB) & 7.688 & 5 & 0.174 & 82.57\% \\
+$H_0$(SHOES) & 9.385 & 5 & 0.095 & 90.53\% \\
\hline
\end{tabular}
\end{center}
\caption{
The $\chi^2$, degrees of freedom (dof), one-tailed p-values (p), and confidence level for DESI 2024 data only for anisotropic BAO observations.
}
\label{table:DESI_sigma_avg}
\end{table*}

\begin{table*}
\begin{center}
DESI24+Pl18(standard) \\
\begin{tabular}{ |c|c|c|c|c| }
\hline
Data & $\chi^2$ ($A_1\&A_2\&A_3$) & dof ($A_1\&A_2\&A_3$) & p ($A_1\&A_2\&A_3$) & CL ($A_1\&A_2\&A_3$) \\
\hline
+$H_0$(tRGB) & 13.739 & 12 & 0.318 & 68.23\% \\
+$H_0$(SHOES) & 20.909 & 12 & 0.052 & 94.83\% \\
\hline
\end{tabular} \\
DESI24+Pl18(standard+$\Sigma m_{\nu}$+$N_{\rm eff}$) \\
\begin{tabular}{ |c|c|c|c|c| }
\hline
Data & $\chi^2$ ($A_1\&A_2\&A_3$) & dof ($A_1\&A_2\&A_3$) & p ($A_1\&A_2\&A_3$) & CL ($A_1\&A_2\&A_3$) \\
\hline
+$H_0$(tRGB) & 13.739 & 12 & 0.318 & 68.23\% \\
+$H_0$(SHOES) & 17.280 & 12 & 0.139 & 86.06\% \\
\hline
\end{tabular} \\
DESI24+Pl18(standard) \\
\begin{tabular}{ |c|c|c|c|c| }
\hline
Data & $\chi^2$ ($B\&A_3$) & dof ($B\&A_3$) & p ($B\&A_3$) & CL ($B\&A_3$) \\
\hline
+$H_0$(tRGB) & 7.718 & 7 & 0.358 & 64.18\% \\
+$H_0$(SHOES) & 13.331 & 7 & 0.064 & 93.56\% \\
\hline
\end{tabular} \\
DESI24+Pl18(standard+$\Sigma m_{\nu}$+$N_{\rm eff}$) \\
\begin{tabular}{ |c|c|c|c|c| }
\hline
Data & $\chi^2$ ($B\&A_3$) & dof ($B\&A_3$) & p ($B\&A_3$) & CL ($B\&A_3$) \\
\hline
+$H_0$(tRGB) & 7.865 & 7 & 0.345 & 65.54\% \\
+$H_0$(SHOES) & 12.197 & 7 & 0.094 & 90.57\% \\
\hline
\end{tabular}
\end{center}
\caption{
The $\chi^2$, degrees of freedom (dof), one-tailed p-values (p), and confidence level for DESI 2024 data for the full set of BAO data (anisotropic and isotropic data combined).
}
\label{table:DESI_sigma_avg_full}
\end{table*}

\begin{table*}
\begin{center}
DESI24 (excluding $z_{\rm eff}=0.51$)+Pl18(standard) \\
\begin{tabular}{ |c|c|c|c|c| }
\hline
Data & $\chi^2$ ($A_1\&A_2\&A_3$) & dof ($A_1\&A_2\&A_3$) & p ($A_1\&A_2\&A_3$) & CL ($A_1\&A_2\&A_3$) \\
\hline
+$H_0$(tRGB) & 9.505 & 12 & 0.659 & 34.07\% \\
+$H_0$(SHOES) & 18.155 & 12 & 0.111 & 88.89\% \\
\hline
\end{tabular} \\
DESI24 (excluding $z_{\rm eff}=0.51$)+Pl18(standard+$\Sigma m_{\nu}$+$N_{\rm eff}$) \\
\begin{tabular}{ |c|c|c|c|c| }
\hline
Data & $\chi^2$ ($A_1\&A_2\&A_3$) & dof ($A_1\&A_2\&A_3$) & p ($A_1\&A_2\&A_3$) & CL ($A_1\&A_2\&A_3$) \\
\hline
+$H_0$(tRGB) & 9.293 & 12 & 0.678 & 32.23\% \\
+$H_0$(SHOES) & 13.997 & 12 & 0.301 & 69.91\% \\
\hline
\end{tabular} \\
DESI24 (excluding $z_{\rm eff}=0.51$)+Pl18(standard) \\
\begin{tabular}{ |c|c|c|c|c| }
\hline
Data & $\chi^2$ ($B\&A_3$) & dof ($B\&A_3$) & p ($B\&A_3$) & CL ($B\&A_3$) \\
\hline
+$H_0$(tRGB) & 3.333 & 7 & 0.853 & 14.74\% \\
+$H_0$(SHOES) & 9.097 & 7 & 0.246 & 75.42\% \\
\hline
\end{tabular} \\
DESI24 (excluding $z_{\rm eff}=0.51$)+Pl18(standard+$\Sigma m_{\nu}$+$N_{\rm eff}$) \\
\begin{tabular}{ |c|c|c|c|c| }
\hline
Data & $\chi^2$ ($B\&A_3$) & dof ($B\&A_3$) & p ($B\&A_3$) & CL ($B\&A_3$) \\
\hline
+$H_0$(tRGB) & 3.333 & 7 & 0.853 & 14.74\% \\
+$H_0$(SHOES) & 7.865 & 7 & 0.345 & 65.54\% \\
\hline
\end{tabular}
\end{center}
\caption{
The $\chi^2$, degrees of freedom (dof), one-tailed p-values (p), and confidence level for DESI 2024 data (anisotropic and isotropic data combined) excluding the anisotropic data at $z_{\rm eff}=0.51$.
}
\label{table:DESI_sigma_avg_full_Again}
\end{table*}

\begin{table*}
\begin{center}
non-DESI BAO+Pl18(standard) \\
\begin{tabular}{ |c|c|c|c|c| }
\hline
Data & $\chi^2$ ($A_1\&A_2$) & dof ($A_1\&A_2$) & p ($A_1\&A_2$) & CL ($A_1\&A_2$) \\
\hline
+$H_0$(tRGB) & 11.881 & 10 & 0.293 & 70.69\% \\
+$H_0$(SHOES) & 22.801 & 10 & 0.012 & 98.85\% \\
\hline
\end{tabular} \\
non-DESI BAO+Pl18(standard+$\Sigma m_{\nu}$+$N_{\rm eff}$) \\
\begin{tabular}{ |c|c|c|c|c| }
\hline
Data & $\chi^2$ ($A_1\&A_2$) & dof ($A_1\&A_2$) & p ($A_1\&A_2$) & CL ($A_1\&A_2$) \\
\hline
+$H_0$(tRGB) & 11.449 & 10 & 0.324 & 67.64\% \\
+$H_0$(SHOES) & 17.161 & 10 & 0.071 & 92.91\% \\
\hline
\end{tabular} \\
non-DESI BAO+Pl18(standard) \\
\begin{tabular}{ |c|c|c|c|c| }
\hline
Data & $\chi^2$ ($B$) & dof ($B$) & p ($B$) & CL ($B$) \\
\hline
+$H_0$(tRGB) & 9.113 & 5 & 0.105 & 89.53\% \\
+$H_0$(SHOES) & 9.660 & 5 & 0.085 & 91.46\% \\
\hline
\end{tabular} \\
non-DESI BAO+Pl18(standard+$\Sigma m_{\nu}$+$N_{\rm eff}$) \\
\begin{tabular}{ |c|c|c|c|c| }
\hline
Data & $\chi^2$ ($B$) & dof ($B$) & p ($B$) & CL ($B$) \\
\hline
+$H_0$(tRGB) & 9.113 & 5 & 0.105 & 89.53\% \\
+$H_0$(SHOES) & 9.800 & 5 & 0.081 & 91.89\% \\
\hline
\end{tabular}
\end{center}
\caption{
The $\chi^2$, degrees of freedom (dof), one-tailed p-values (p), and confidence level for non-DESI BAO data.
}
\label{table:nondesi_sigma_avg}
\end{table*}

\section{Results}
\label{sec-result}

Putting $\tilde{D}_M$ from Table~\ref{table:DESI24_DM_DH_tilde}, $\omega_{\rm m0}$ and $r_d$ from Table~\ref{table:wm0_rd}, $h$ from Table~\ref{table:h_from_H0_data} in to Eq.~\eqref{eq:diagnostic_A1}, we compute $A_1$ for DESI 2024 BAO data. We plot these values in Fig.~\ref{fig:DESI_A1}. For a fixed $H_0$ value, we see the deviations of $A_1$ from $1$ are almost independent of early-time physics. For tRGB $H_0$, the deviations are less than 1$\sigma$ to around 1$\sigma$. For SHOES $H_0$ the deviations are less than 1$\sigma$ to around 2$\sigma$ except at $z_{\rm eff}=0.51$, where it is around 3$\sigma$. This is because the CMB constraints on the deviations are tighter compared to the constraints from local distance measures. The higher the $H_0$ value, (moderately) higher the deviations at most of the effective redshift points.

We plot diagnostic variable $A_2$ in Fig.~\ref{fig:DESI_A2} for DESI 2024 data. Here, also we see a very weak dependence of deviations on early physics. The deviations are up to 1$\sigma$ for tRGB $H_0$, except for $z_{\rm eff}=0.51$ where it is around little less than 2$\sigma$. For SHOES $H_0$ deviations are up to 1$\sigma$, except for $z_{\rm eff}=0.93$ where it is around 2$\sigma$. The deviations of $A_2$ are slightly larger for the higher value of $H_0$ compared to the lower value of $H_0$ but not very significantly larger.

Similarly, we plot $A_1$ and $A_2$ in Figs.~\ref{fig:nondesi_A1} and~\ref{fig:nondesi_A2} respectively for other non-DESI BAO data. The same conclusions are applicable here as in the case of DESI 2024 BAO data for $A_1$ and $A_2$ respectively (only individual deviations are slightly different).

By using the values of $F_{\rm AP}(z)$ from Table~\ref{table:DESI_24_FAP_data}, $\omega_{\rm m0}$ from Table~\ref{table:wm0_rd}, $h$ from Table~\ref{table:h_from_H0_data} and using Eq.~\eqref{eq:defn_B}, we compute diagnostic variable $B(z)$ at each effective redshift. We plot $B$ and $\Delta B$ in Fig.~\ref{fig:B_check_DESI}. We see the deviations of $B(z)$ from the value $1$ i.e. the deviations from the $\Lambda$CDM model are almost independent of the early-time physics (irrespective of whether we consider standard early-time physics or the early cosmological model with the varying sum of neutrino masses $\Sigma m_{\nu}$ and effective number of relativistic species $N_{\rm eff}$). The deviations are slightly different for different $H_0$ but not very significant. The deviations are well less than 1$\sigma$ to little above 1$\sigma$, except at $z_{\rm eff}=0.51$, where it is little above 2$\sigma$.

Similarly, we plot $B$ for other non-DESI BAO data in Fig.~\ref{fig:B_non_DESI_BAO}. The same conclusion is applicable here too (only slightly different deviations at each effective redshift). The deviations are up to around 1$\sigma$ except at $z_{\rm eff}=0.698$, where it is a little above 2$\sigma$.

To check how the individual deviations are for $A_3$ corresponding to DESI24 data, we plot these deviations in Fig.~\ref{fig:desi_plot_A3}. Here also we see the individual deviations are almost independent of early-physics. The deviations have little dependence on $H_0$. For tRGB $H_0$, the deviations are well within 1$\sigma$. For SHOES $H_0$ the deviations are a little above 1$\sigma$ to around 2$\sigma$. We find the correlation coefficients between them are around 0.2 to 0.4, which is quite large.

The quantification of each deviation at each effective redshift is not very useful to make any conclusion. For this, we do a goodness-of-fit hypothesis rejection test using $\chi^2$ statistics, obtained from the combination of all the deviations at all effective redshifts. To do so, we define $\chi^2$ as

\begin{equation}
\chi^2 = v^T C^{-1}v ,
\label{eq:average_sigma}
\end{equation}

\noindent
where $v$ is the vector consisting of all these deviations and $C$ is the corresponding covariance matrix. In the above equation, the chi-square is defined in such a way that it takes into account the correlation between each deviation, which we can not see in the corresponding plots, we have discussed so far. Note that, because of the presence of the non-zero covariances, the non-diagonal elements are non-zero, in general, and the square root of the diagonal elements are the standard deviations at each effective redshift, which are plotted by error bars in the above figures. From this $\chi^2$ value, the corresponding degrees of freedom (denoted as 'dof') and using Gaussian distribution, we compute one-tailed p-values (denoted as 'p') and consequently, we compute the confidence interval (denoted as 'CL') in percentage to reject the hypothesis.

We list all these values in Table~\ref{table:DESI_sigma_avg} for anisotropic DESI 2024 BAO data using Eq.~\eqref{eq:average_sigma}. We can see that the rejection of the flat $\Lambda$CDM model is only around 80\% for tRGB values of $H_0$ with combinations of data involving DESI24 and CMB. These are weak evidence against the flat $\Lambda$CDM model. For the SHOES values of $H_0$, the confidence intervals are slightly higher around 90-97\%. Still, these are also not strong enough. Further, we find that the confidence intervals have no significant dependence on the early physics. All these conclusions are almost similar between combined $A_1\&A_2$ and $B$.

Note that, if we consider only diagnostic $A_1$, we miss information from $\tilde{D}_H$ from BAO data. Similarly, if we consider only diagnostic $A_2$, we miss information from $\tilde{D}_M$. Also, we compute incorrect error estimation because we miss the cross-covariances between $\tilde{D}_M$ and $\tilde{D}_H$ from BAO data. But, when we consider $A_1$ and $A_2$ together, these issues are not present. Note that, a similar conclusion about evidence of dynamical dark energy is obtained from combined $A_1\&A_2$ and $B$. This can be seen in Table~\ref{table:DESI_sigma_avg} and the next tables. Thus, $B$ can almost be equivalently used instead of combined $A_1\&A_2$ when we study the evidence for dynamical dark energy.

Also note that, alongside the correlation between $\tilde{D}_M$ and $\tilde{D}_H$, in principle, there should also be correlations between each $B$ (or in $A_1$ and $A_2$ along with all their self and cross-covariances) between each effective redshift, because each $B$ shares common parameter $\alpha$ (and other parameters for $A_1$ and $A_2$). However, this correlations are weak for $B$ (of the order $10^{-2}$ to little less than $10^{-1}$), but this correlations are not weak in $A_1$ and $A_2$ (of the order $10^{-1}$ to little less than $1$). This is because the correlations between $\tilde{D}_M$ and $\tilde{D}_H$ were already taken into account in B through $F_{\rm AP}$ but not in $A_1$ and $A_2$ and also, $A_1$ and $A_2$ involve extra parameters. This is a reason that diagnostic $B$ is better than the individual diagnostics $A_1$ and $A_2$ when we plot individual deviations or individual confidence intervals, but again these are almost equivalent when we combine $A_1$ and $A_2$ by including all the self and cross covariances between $A_1$ and $A_2$ for combined confidence interval.

DESI 2024 observations also provide isotropic BAO data. So, to get fully constrained results, we include these data to compute confidence intervals. These are listed in Table~\ref{table:DESI_sigma_avg_full}. The inclusion of isotropic BAO data does not change the confidence intervals significantly to reject the standard $\Lambda$CDM model for the late time evolution. One interesting fact to notice here is that the inclusion of isotropic BAO data makes the differences between the combinations $A_1\&A_2\&A_3$ and $B\&A_3$ smaller compared to the differences between the combinations $A_1\&A_2$ and $B$. This proves the equivalence between $A_1\&A_2\&A_3$ and $B\&A_3$.

Next, in Table~\ref{table:DESI_sigma_avg_full_Again}, we exclude the anisotropic DESI 2024 data at $z_{\rm eff}=0.51$ to see if this exclusion significantly changes the results. We find this exclusion does not change the results significantly (changes are only around 0.1$\sigma$ to 0.2$\sigma$).

In summary, we get no significant evidence of dynamical dark energy when we combine the DESI24 data, the CMB observations from the Planck 2018 mission, and the $H_0$ values from the local measurements.

Now, we turn our attention to the other non-DESI BAO data combined with the same CMB and $H_0$ data. Similar to previous tables, we list corresponding $\chi^2$ values, number of degrees of freedom, one-tailed p-values, and confidence interval in Table~\ref{table:nondesi_sigma_avg}. We find the confidence intervals are around 68-93\% to reject $\Lambda$CDM model except for non-DESI BAO+Pl18(standard)+$H_0$(SHOES) for which it is around 99\%. This means evidence of dynamical dark energy is weak, except for non-DESI BAO+Pl18(standard)+$H_0$(SHOES) combination of data for which it is moderate but not very strong enough. Here, also, we see the confidence intervals are almost independent of the early-time physics, except the fact that the confidence interval for non-DESI BAO+Pl18(standard)+$H_0$(SHOES) is larger compared to the one for non-DESI BAO+Pl18(standard+$\Sigma m_{\nu}+N_{\rm eff}$)+$H_0$(SHOES).

So, with the analysis, presented in this study, with the diagnostic null test, we find no strong evidence of dynamical dark energy both for DESI 2024 and other BAO data combined with CMB and $H_0$ observations.

\section{Conclusion}
\label{sec-conclusion}

We have proposed new diagnostics for the null tests for the evidence of dynamical dark energy. These diagnostics measure the deviation from the $\Lambda$CDM model for the late-time background evolution of the Universe. These diagnostics are independent of any late-time dark energy model or parametrization, however, the derivation of these diagnostics depends on the basic cosmological assumptions that the background geometry of the Universe is described by the flat FLRW metric. This kind of null test is crucial to avoid any model-dependent bias, any dependence on priors of the dark energy model parameters, and the presence of any degeneracies in the estimations of the dark energy parameters.

We have defined the diagnostics by quantities $A_1$, $A_2$, $B$, and $A_3$ as functions of redshift. For the $\Lambda$CDM model of the late-time background dynamics of the Universe, the values of these quantities are unity at any redshift. The deviations in the observed values of these diagnostics, obtained from any observations, from $1$ correspond to the deviations from the $\Lambda$CDM model. Thus, any significant deviations from $1$ correspond to the evidence for the dynamical dark energy.

With these diagnostics, we study the deviations from the $\Lambda$CDM model with the combinations of the latest BAO measurements from DESI 2024 Data Release 1 (DR1) data, CMB data from Planck 2018 mission (TT,TE,EE+lowE+lensing), and local measurements of $H_0$ (both from tRGB and SHOES observations). We also consider other BAO observations mostly from SDSS IV data in our analysis.

When DESI 2024 and the CMB observations are combined with the $H_0$ measurement of tRGB observations, the evidence of dynamical dark energy is weak. When we replace the tRGB $H_0$ value with SHOES $H_0$ value, we find that the evidence of dynamical dark energy is comparatively stronger but not strong enough to reject the $\Lambda$CDM model. The exclusion of the DESI 2024 data at effective redshift 0.51 makes the evidence further weaker.

When we consider other BAO data instead of DESI 2024 data, we see similar results and find similar conclusions compared to the DESI 2024 data.

We find that the evidence of dynamical dark energy is almost independent of any early-time physics for all the combinations of data.

\acknowledgments
The author is supported by the South African Radio Astronomy Observatory and National Research Foundation (Grant No. 75415). The author would like to thank Roy Maartens for useful comments on the draft.

\bibliographystyle{JHEP}
\bibliography{references}





\appendix

\section{Analytical expression of comoving distance in $\Lambda$CDM model}
\label{sec-los_Lcdm}

In a flat FLRW cosmological background, the transverse comoving distance is equal to the line of sight comoving distance and in the $\Lambda$CDM model it is given as (by putting Eq.~\eqref{eq:E_LCDM_late} in Eq.~\eqref{eq:defn_DM})

\begin{equation}
D_M^{\Lambda CDM} (z) = \frac{c}{H_0} I(z),
\label{eq:DM_Lcdm_late_2}
\end{equation}

\noindent
where $I$ is given as

\begin{equation}
I(z) = \int_{0}^{z} \frac{d\tilde{z}}{\sqrt{\Omega_{\rm m0}(1+\tilde{z})^3+1-\Omega_{\rm m0}}} = \beta \int_{0}^{z} \left[ 1+\alpha(1+\tilde{z})^3 \right]^{-\frac{1}{2}} d\tilde{z} ~ ,
\label{eq:los_Lcdm_I}
\end{equation}

\noindent
where $\beta$ is given as

\begin{equation}
\beta = \frac{1}{\sqrt{1-\Omega_{\rm m0}}} .
\label{eq:beta}
\end{equation}

\noindent
Let us change the integration variable from $\tilde{z}$ to $x$ such that

\begin{equation}
x = (1+\tilde{z})^3.
\label{eq:DM_LCDM_late_2}
\end{equation}

\noindent
Using this variable, $I(z)$ in Eq.~\eqref{eq:los_Lcdm_I} becomes

\begin{equation}
I(z) = \frac{\beta}{3} \int_{\gamma(z=0)}^{\gamma(z)} x^{-\frac{2}{3}} (1+\alpha x)^{-\frac{1}{2}} dx,
\label{eq:los_Lcdm_I_3}
\end{equation}

\noindent
where $\gamma$ is given as

\begin{equation}
\gamma (z) = (1+z)^3.
\label{eq:los_Lcdm_gamma}
\end{equation}

\noindent
Eq.~\eqref{eq:los_Lcdm_I_3} can be rewritten as

\begin{equation}
I(z) = \frac{\beta}{3} \left[ \tilde{I}(z)-\tilde{I}(z=0) \right],
\label{eq:los_Lcdm_I_4}
\end{equation}

\noindent
where $\tilde{I}$ is given as

\begin{equation}
\tilde{I} (z) = \int_{0}^{\gamma(z)} x^{-\frac{2}{3}} (1+\alpha x)^{-\frac{1}{2}} dx.
\label{eq:los_Lcdm_tildeI}
\end{equation}

\noindent
Let us again change the variable from $x$ to $y$ such that

\begin{equation}
y = \frac{x}{\gamma (z)}.
\label{eq:los_Lcdm_y}
\end{equation}

\noindent
With the changed variable, Eq.~\eqref{eq:los_Lcdm_tildeI} can be rewritten as

\begin{eqnarray}
\tilde{I} (z) &=& (1+z) f(z),
\label{eq:los_Lcdm_tildeI_2} \\
f (z) &=& \int_{0}^{1} y^{-\frac{2}{3}} \left[1-b(z)y\right]^{-\frac{1}{2}} dy,
\label{eq:los_Lcdm_tildeI_3_main_f}
\end{eqnarray}

\noindent
where we have used the expression of $\gamma$ from Eq.~\eqref{eq:los_Lcdm_gamma}. The expression of $b$ is given as

\begin{equation}
b(z) = -\alpha \gamma(z) = -\alpha(1+z)^3 .
\label{eq:los_b}
\end{equation}

\noindent
The expression $f$ in Eq.~\eqref{eq:los_Lcdm_tildeI_3_main_f} can be rewritten as

\begin{eqnarray}
f (z) &=& \int_{0}^{1} y^{a_2-1} (1-y)^{a_3-a_2-1} \left[1-b(z)y\right]^{-a_1} dy,
\label{eq:los_Lcdm_tildeI_4_main_f_2} \\
a_1 &=& \frac{1}{2}, ~~~ a_2 = \frac{1}{3}, ~~~ a_3 = \frac{4}{3}.
\label{eq:los_a1_a2_a3}
\end{eqnarray}

\noindent
The integration $f$ in Eq.~\eqref{eq:los_Lcdm_tildeI_4_main_f_2} is the standard integral form of the hypergeometric function given as

\begin{eqnarray}
\int_{0}^{1} y^{a_2-1} (1-y)^{a_3-a_2-1} \left[1-by\right]^{-a_1} dy &=& \frac{\Gamma(a_2)\Gamma(a_3-a_2)}{\Gamma(a_3)} \, _2F_1\left(a_1,a_2;a_3;b\right) \nonumber\\
&=& \frac{\Gamma(a_2)\Gamma(a_3-a_2)}{\Gamma(a_3)} \, _2F_1\left(a_2,a_1;a_3;b\right) ,
\label{eq:hyp_integral}
\end{eqnarray}

\noindent
where $\Gamma$ stands for the usual Gamma function. Using the integration result of $f$ from Eq.~\eqref{eq:hyp_integral}, using the values of $a_1$, $a_2$, and $a_3$ from Eq.~\eqref{eq:los_a1_a2_a3}, and using the expression of $b$ from Eq.~\eqref{eq:los_b}, we get the analytical form of $\tilde{I}$ in Eq.~\eqref{eq:los_Lcdm_tildeI_2} given as

\begin{equation}
\tilde{I}(z) = 3 (1+z) \, _2F_1\left[1/3,1/2;4/3;-\alpha(1+z)^3\right].
\label{eq:analytic_tilde_I}
\end{equation}

\noindent
Using $\tilde{I}(z)$ from Eq.~\eqref{eq:analytic_tilde_I} in to Eq.~\eqref{eq:los_Lcdm_I_4}, we get the expression of $I$ given as

\begin{equation}
I(z) = \beta \left( (1+z) \, _2F_1\left[1/3,1/2;4/3;-\alpha(1+z)^3\right] - \, _2F_1\left[1/3,1/2;4/3;-\alpha\right] \right) = \beta F(z) .
\label{eq:analytic_I}
\end{equation}

\noindent
Finally, putting expressions of $\beta$ from Eq.~\eqref{eq:beta} and $I$ from Eq.~\eqref{eq:analytic_I} in Eq.~\eqref{eq:DM_Lcdm_late_2}, we get analytical expression of $D_M^{\Lambda CDM}$ given as

\begin{equation}
D_M^{\Lambda CDM} (z) = \frac{c F(z)}{H_0 \sqrt{1-\Omega_{\rm m0}} } .
\label{eq:DM_Lcdm_late_3_analytic}
\end{equation}

\noindent
Hence the result in Eq.~\eqref{eq:DM_Lcdm_late_3_analytic} is the same as in Eq.~\eqref{eq:DM_LCDM_late} in the main text.

\section{Approximate sound horizon at the early time with standard early time physics}
\label{sec-appx_sound_horzon_early_standard}

The sound horizon $r_s$ is defined as

\begin{equation}
r_s (z) = \int_z^{\infty} \frac{c_s(\tilde{z}) d\tilde{z}}{H(\tilde{z})} = \int_0^{\frac{1}{1+z}} \frac{c_s(a) da}{a^2H(a)},
\label{eq:sound_horizon_general}
\end{equation}

\noindent
where $c_s$ is the sound speed, $a$ is the cosmic scale factor and it is related to redshift as $a=\frac{1}{1+z}$. For $z \lesssim z_*$, sound speed $c_s$ can be approximated as

\begin{equation}
c_s (a) = \frac{c}{\sqrt{3\left(1+\frac{3}{4}\frac{\rho_b(a)}{\rho_{\gamma}(a)}\right)}} = \frac{c}{\sqrt{3\left(1+\frac{3}{4}\frac{\omega_{\rm b0}}{\omega_{\rm \gamma 0}}a\right)}},
\label{eq:sound_speed_appx}
\end{equation}

\noindent
where $\rho_b$ and $\rho_{\gamma}$ are the baryon and photon energy densities respectively. In the second equality in Eq.~\eqref{eq:sound_speed_appx}, $\omega_{\rm \gamma 0}$ is defined as

\begin{equation}
\omega_{\rm \gamma 0} = \Omega_{\rm \gamma 0} h^2 ,
\label{eq:defn_wg0}
\end{equation}

\noindent
where $\Omega_{\rm \gamma 0}$ is the photon energy density parameter at present.

The sound speed in Eq.~\eqref{eq:sound_speed_appx} can be rewritten as

\begin{equation}
c_s (a) = \frac{c\sqrt{\tilde{\alpha}}}{\sqrt{3}} \frac{1}{\sqrt{a+\tilde{\alpha}}},
\label{eq:sound_speed_appx_2}
\end{equation}

\noindent
where $\tilde{\alpha}$ is given as

\begin{equation}
\tilde{\alpha} = \frac{\tilde{\beta}}{\omega_{\rm b0}}.
\label{eq:tilde_alpha}
\end{equation}

\noindent
In Eq.~\eqref{eq:tilde_alpha}, $\tilde{\beta}$ is given as \citep{Chen:2018dbv,Zhai:2018vmm}

\begin{equation}
\tilde{\beta} = \frac{4\omega_{\rm \gamma 0}}{3} = \frac{\theta^{4}}{31500},
\label{eq:tilde_beta}
\end{equation}

\noindent
where $\theta$ is given as \citep{Chen:2018dbv,Zhai:2018vmm}

\begin{eqnarray}
\theta &=& \frac{T_{\rm CMB}}{2.7~\rm Kelvin},
\label{eq:cmb_theta} \\
T_{\rm CMB} &=& 2.7255~{\rm Kelvin}.
\label{eq:temp_CMB}
\end{eqnarray}

\noindent
Putting the value of present CMB temperature $T_{\rm CMB}$ from Eq.~\eqref{eq:temp_CMB} and using Eqs.~\eqref{eq:cmb_theta} and~\eqref{eq:tilde_beta}, we get approximate value of $\tilde{\beta}$ as

\begin{equation}
\tilde{\beta} \approx 3.29624 \times 10^{-5} ~.
\end{equation}

\noindent
By putting Eq.~\eqref{eq:sound_speed_appx_2} in Eq.~\eqref{eq:sound_horizon_general} and using the definition $H(z)=H_0E(z)$, we get an expression for the sound horizon given as

\begin{equation}
r_s (z) = \frac{c\sqrt{\tilde{\alpha}}}{H_0\sqrt{3}} \int_0^{\frac{1}{1+z}} \frac{da}{a^2E(a)\sqrt{a+\tilde{\alpha}}} .
\label{eq:sound_horizon_2}
\end{equation}

\noindent
In a flat FLRW background, with standard early-time physics, the normalized Hubble parameter can be approximated as

\begin{equation}
E_{\rm standard}^{\rm early-time}(z) \approx \sqrt{\Omega_{\rm m0}(1+z)^3+\Omega_{\rm r0}(1+z)^4},
\label{eq:E_standard_early}
\end{equation}

\noindent
where $\Omega_{\rm r0}$ is the present value of the radiation energy density parameter. Putting Eq.~\eqref{eq:E_standard_early} in Eq.~\eqref{eq:sound_horizon_2} and using the expression in Eq.~\eqref{eq:H0_by_c}, the integral in Eq.~\eqref{eq:sound_horizon_2} becomes

\begin{equation}
r_s (z) = \frac{3000 \sqrt{\tilde{\alpha}} ~ {\rm Mpc}}{\sqrt{3 \omega_{\rm m0}}} \int_0^{\frac{1}{1+z}} \frac{da}{\sqrt{(a+\tilde{\alpha})(a+\tilde{\gamma})}} .
\label{eq:sound_horizon_3}
\end{equation}

\noindent
In Eq.~\eqref{eq:sound_horizon_3}, $\tilde{\gamma}$ is given as

\begin{equation}
\tilde{\gamma} = \frac{\Omega_{\rm r0}}{\Omega_{\rm m0}} = \frac{1}{1+z_{\rm eq}^{\rm m-r}},
\label{eq:gamma_tilde}
\end{equation}

\noindent
where $z_{\rm eq}^{\rm m-r}$ is the redshift of radiation and matter equality. It can be written approximately as \citep{Chen:2018dbv,Zhai:2018vmm}

\begin{equation}
z_{\rm eq}^{\rm m-r} = 25000~\omega_{\rm m0} \theta^{-4} \approx 24077.44059~\omega_{\rm m0}.
\label{eq:z_eq_rad_mat}
\end{equation}

\noindent
In the last (approximate) equality of Eq.~\eqref{eq:z_eq_rad_mat}, we have used Eq.~\eqref{eq:temp_CMB} and Eq.~\eqref{eq:cmb_theta}. Eq.~\eqref{eq:sound_horizon_3} can be rewritten as

\begin{equation}
r_s (z) = \tilde{\delta} \int_0^{\frac{1}{1+z}} \frac{da}{\sqrt{(a+\tilde{\chi})^2-\xi^2}} ,
\label{eq:sound_horizon_4}
\end{equation}

\noindent
where $\tilde{\delta}$, $\tilde{\chi}$, and $\xi$ are given as

\begin{eqnarray}
\tilde{\delta} &=& \frac{3000 \sqrt{\tilde{\beta}} ~ {\rm Mpc}}{\sqrt{3 \omega_{\rm m0} \omega_{\rm b0}}},
\label{eq:delta_tilde} \\
\tilde{\chi} &=& \frac{\tilde{\alpha}+\tilde{\gamma}}{2},
\label{eq:tilde_chi} \\
\xi &=& \frac{\tilde{\alpha}-\tilde{\gamma}}{2} ,
\label{eq:xi}
\end{eqnarray}

\noindent
respectively. Now, let us change the integration variable from $a$ to $\tilde{a}$ in Eq.~\eqref{eq:sound_horizon_4} such that

\begin{equation}
\tilde{a} = a+\tilde{\chi},
\label{eq:tilde_a}
\end{equation}

\noindent
and with this changed variable, Eq.~\eqref{eq:sound_horizon_4} can be rewritten as

\begin{equation}
r_s (z) = \tilde{\delta} \int_{\tilde{\chi}}^{\frac{1}{1+z}+\tilde{\chi}} \frac{d \tilde{a}}{\sqrt{\tilde{a}^2-\xi^2}} .
\label{eq:sound_horizon_5}
\end{equation}

\noindent
We now use the integration result given as

\begin{eqnarray}
\int \frac{d \tilde{a}}{\sqrt{\tilde{a}^2-\xi^2}} = {\rm log} \left| \tilde{a}+\sqrt{\tilde{a}^2-\xi^2} \right| + {\rm constant}.
\label{eq:integration_2}
\end{eqnarray}

\noindent
Note that, here all the variables are positive definite. So, we can neglect the absolute sign in Eq.~\eqref{eq:integration_2}. Using this equation and using Eqs.~\eqref{eq:tilde_chi} and~\eqref{eq:xi}, we finally get an analytical expression for the approximate sound horizon given as

\begin{equation}
r_s (z) = \tilde{\delta} ~ {\rm log} \left( \frac{{2}+(1+z)(\tilde{\alpha}+\tilde{\gamma})+2\big[(1+z)^2{\tilde{\alpha}\tilde{\gamma}+ (1+z)({\tilde{\alpha}+\tilde{\gamma}})+1}\big]^{1/2}}{(1+z)\big[{\tilde{\alpha}+\tilde{\gamma}}+2({\tilde{\alpha}\tilde{\gamma}})^{1/2}\big]} \right) ,
\label{eq:sound_horizon_6}
\end{equation}

\noindent
From Eq.~\eqref{eq:sound_horizon_6} and the previous equations, we can see that, for the standard early time physics, the sound horizon depends only on $\omega_{\rm m0}$ and $\omega_{\rm b0}$ parameters.

\section{Approximate redshift of photon decoupling with standard early time physics}
\label{sec-appx_photon_decoupling_z_standard}

With the standard early-time physics, the redshift of photon decoupling $z_*$ can be approximated to have an analytical expression given as \citep{Chen:2018dbv,Zhai:2018vmm}

\begin{equation}
z_* = 1048 \left( 1+0.00124 \omega_{\rm b0}^{-0.738} \right) \left( 1+g_1 \omega_{\rm m0}^{g_2} \right),
\label{eq:z_photon_decoupling}
\end{equation}

\noindent
where $g_1$ and $g_2$ are given as

\begin{eqnarray}
g_1 &=& \frac{ 0.0783 \omega_{\rm b0}^{-0.238} }{ 1+39.5 \omega_{\rm b0}^{0.763} },
\label{eq:z_cmb_g1} \\
g_2 &=& \frac{0.560}{ 1+21.1 \omega_{\rm b0}^{1.81} },
\label{eq:z_cmb_g_2}
\end{eqnarray}

\noindent
respectively. Putting Eq.~\eqref{eq:z_photon_decoupling} in Eq.~\eqref{eq:sound_horizon_6}, we get sound horizon at photon decoupling redshift i.e. $r_s(z_*)$ for standard early-time physics.

\section{Approximate redshift of baryon drag epoch with standard early time physics}
\label{sec-appx_baryon_drag_z_standard}

The redshift of the baryon drag epoch $z_d$ has an approximate expression given as \citep{Hu:1995en}:

\begin{equation}
z_d = \frac{1345\, \omega_{\rm m0}^{0.251}\big(1+b_1\,\omega_{\rm b0}^{b_2}\big)}{1+0.659\,\omega_{\rm m0}^{0.828}} ,
\label{eq:z_baryon_drag}
\end{equation}

\noindent
for standard early-time physics, where we have

\begin{eqnarray}
b_1 &=& 0.313\, \omega_{\rm m0}^{-0.419}\left(1+0.607\,\omega_{\rm m0}^{0.674}\right) ,
\label{eq:z_d_b1} \\
b_2 &=& 0.238\,\omega_{\rm m0}^{0.223} .
\label{eq:z_d_b2}
\end{eqnarray}

\noindent
Putting Eq.~\eqref{eq:z_baryon_drag} in Eq.~\eqref{eq:sound_horizon_6}, we get sound horizon at baryon drag epoch i.e. $r_d$.

\section{CMB distance priors}
\label{sec-cmb_dist_prior_again}

We consider Planck 2018 results for TT,TE,EE+lowE+lensing for which we have \citep{Zhai:2018vmm}

\begin{equation}
v (\text{Pl18(standard)}) =
\begin{bmatrix} 
R \\
l_A \\
\omega_{\rm b0}
\end{bmatrix}
=
\begin{bmatrix} 
1.74963 \\
301.80845 \\
0.02237
\end{bmatrix}
.
\end{equation}

\noindent
The corresponding covariance matrix is given as

\begin{equation}
C (\text{Pl18(standard)}) = 10^{-8} \times
\begin{bmatrix} 
1598.9554 & 17112.007 & -36.311179 \\
17112.007 & 811208.45 & -494.79813 \\
-36.311179 & -494.79813 & 2.1242182
\end{bmatrix}
.
\end{equation}

\noindent
From the above two equations, we get a corresponding mean vector for $Q$ and $\omega_{\rm b0}$ given as

\begin{equation}
v_2 (\text{Pl18(standard)}) =
\begin{bmatrix}
Q \\
\omega_{\rm b0}
\end{bmatrix}
=
\begin{bmatrix}
0.005797 \\
0.02237
\end{bmatrix}
,
\end{equation}

\noindent
and the corresponding covariance matrix is given as

\begin{equation}
C_2 (\text{Pl18(standard)}) = 10^{-10} \times
\begin{bmatrix}
1.56751 & -11.08079 \\
-11.08079 & 212.42182
\end{bmatrix}
.
\end{equation}

Next, from the above two equations and the functional form of $r_s(z_*)$, we get corresponding mean vectors for $\omega_{\rm m0}$ and $\omega_{\rm b0}$ and the corresponding covariance matrix. Finally, from these and using the functional form of $r_d$, we get corresponding mean values of $\omega_{\rm m0}$ and $r_d$, errors in these, and correlations between them. These are listed in the first row of Table~\ref{table:wm0_rd} in the main text.

We also consider a non-standard early-physics cosmology for the same CMB data with two extra degrees of freedom by allowing variations in the sum of neutrino masses $\Sigma m_{\nu}$ and effective number of relativistic species $N_{\rm eff}$. For this case, we follow Zhai et al. (2020) \citep{Zhai:2019nad}, and the mean vector for the CMB distance priors is given as

\begin{equation}
v \left(\text{Pl18(standard+$\Sigma m_{\nu}$+$N_{\rm eff}$)}\right) =
\begin{bmatrix} 
R \\
l_A \\
\omega_{\rm b0} \\
\omega_{\rm c0} \\
N_{\rm eff}
\end{bmatrix}
=
\begin{bmatrix} 
1.7661 \\
301.7293 \\
0.02191 \\
0.1194 \\
2.8979
\end{bmatrix}
,
\end{equation}

\noindent
where $\omega_{\rm c0}$ is the present value of the energy density parameter corresponding to the only cold dark matter such that $\omega_{\rm m0}=\omega_{\rm c0}+\omega_{\rm b0}$. The corresponding covariance matrix is given as

\begin{eqnarray}
&& C (\text{Pl18(standard+$\Sigma m_{\nu}$+$N_{\rm eff}$)}) \nonumber\\
&& = 10^{-8} \times
\begin{bmatrix} 
33483.54 & -44417.15 & -515.03 & -360.42 & -274151.72 \\
-44417.15 & 4245661.67 & 2319.46 & 63326.47 & 4287810.44 \\
-515.03 & 2319.46 & 12.92 & 51.98 & 7273.04 \\
-360.42 & 63326.47 & 51.98 & 1516.28 & 92013.95 \\
-274151.72 & 4287810.44 & 7273.04 & 92013.95 & 7876074.60
\end{bmatrix}
.
\end{eqnarray}

\noindent
From the above two equations, we get the mean vector for $Q$, $\omega_{\rm b0}$, $\omega_{\rm b0}$, and $N_{\rm eff}$ given as

\begin{equation}
v_2 \left(\text{Pl18(standard+$\Sigma m_{\nu}$+$N_{\rm eff}$)}\right) =
\begin{bmatrix} 
Q \\
\omega_{\rm b0} \\
\omega_{\rm m0} \\
N_{\rm eff}
\end{bmatrix}
=
\begin{bmatrix}
0.005853 \\
0.02191 \\
0.1413 \\
2.8979
\end{bmatrix}
,
\end{equation}

\noindent
and the corresponding covariance matrix is given as

\begin{eqnarray}
&& C_2 (\text{Pl18(standard+$\Sigma m_{\nu}$+$N_{\rm eff}$)}) \nonumber\\
&& = 10^{-9} \times
\begin{bmatrix}
3.75 & -17.52 & -41.75 & -9917.81 \\
-17.52 & 129.2 & 649.0 & 72730.4 \\
-41.75 & 649.0 & 16331.60 & 992869.9 \\
-9917.81 & 72730.4 & 992869.9 & 78760746.0
\end{bmatrix}
.
\end{eqnarray}

\noindent
Finally, using a similar procedure, from the above two equations, we get mean values of $\omega_{\rm m0}$ and $r_d$, errors in these, and correlations between them \citep{Zhai:2019nad}. These are listed in the second row of Table~\ref{table:wm0_rd} in the main text.

\section{Brief overview of wCDM and $w_0w_a$CDM parametrizations}
\label{sec-dark_energy_parametrizations}

The equation of state of dark energy $w$ in the wCDM parametrization, and the Chevallier-Polarski-Linder (CPL) parametrization \citep{Chevallier:2000qy,Linder:2002et} ($w_0w_a$CDM parametrization) are given as

\begin{eqnarray}
w(z) &=& w ~~~~~ ({\rm wCDM}),
\label{eq:wCDM} \\
w(z) &=& w_0+w_a \frac{z}{1+z} ~~~~~ (w_0w_a{\rm CDM}),
\label{eq:w0waCDM}
\end{eqnarray}

\noindent
respectively. The standard $\Lambda$CDM model corresponds to $w=-1$ (in wCDM) or $w_0=-1$ and $w_a=0$ (in $w_0w_a$CDM). In general, in any dark energy parametrization, the square of the normalized Hubble parameter has the expression (neglecting radiation) given as

\begin{equation}
E^2(z) = \Omega_{\rm m0}(1+z)^{3}+(1-\Omega_{\rm m0}) \exp \left[ 3 \int_0^z \frac{1+w(\tilde{z})}{1+\tilde{z}} d\tilde{z} \right] .
\label{eq:DE_Esqr}
\end{equation}

\end{document}